\documentclass{emulateapj}

\usepackage{graphics,graphicx,rotating}
\usepackage{natbib}
\usepackage{xspace}
\usepackage{amssymb}
\usepackage{amsmath}
\usepackage{epstopdf}
\usepackage{subfigure}
\usepackage{xcolor}

\shorttitle{Hydrodynamical Simulations of Colliding Jets}
\shortauthors{Molnar et al.}

\newcommand{\simless} 
     {\ensuremath{\lower 3pt\hbox{$\rlap{\raise5pt\hbox{$\char'074$}}\mathchar"7218$}}}
\newcommand{\simgreat}
     {\ensuremath{\lower 3pt\hbox{$\rlap{\raise5pt\hbox{$\char'076$}}\mathchar"7218$}}}

\newcommand{\simgt}{\lower.5ex\hbox{$\; \buildrel > \over \sim \;$}}
\newcommand{\simlt}{\lower.5ex\hbox{$\; \buildrel < \over \sim \;$}}

\newcommand{\DEG}{{$^{\circ}$}}

\newcommand{\KMSEC}{{$\rm km\;s^{-1}$}\xspace}

\newcommand*{\ltsim}{\ {\raise-.75ex\hbox{$\buildrel<\over\sim$}}\ }
\newcommand*{\gtsim}{\ {\raise-.75ex\hbox{$\buildrel>\over\sim$}}\ }
\newcommand*{\proptosim}{\ {\raise-.75ex\hbox{$\buildrel\propto\over\sim$}}\ }

\newcommand{\THREEC}{{3C~75}\xspace}

\newcommand{\bomega}{\pmb{\omega}}

\def\bv{\mbox{\boldmath $v$}\xspace}

\newenvironment{amssidewaysfigure}
  {\begin{sidewaysfigure}\vspace*{0.0\textwidth}\begin{minipage}{\textheight}\centering}
  {\end{minipage}\end{sidewaysfigure}}

\newenvironment{evenamssidewaysfigure}
  {\begin{sidewaysfigure}\vspace*{-0.0\textwidth}\begin{minipage}{\textheight}\centering}
  {\end{minipage}\end{sidewaysfigure}}

\newcommand*{\CHANDRA}{\emph{Chandra}\xspace}
\newcommand*{\XMM}{\emph{XMM-Newton}\xspace}

\newcommand*{\GAMER}{\emph{GAMER}\xspace}
\newcommand*{\FLASH}{\emph{FLASH}\xspace}

\begin{document}

\title{Hydrodynamical Simulations of Colliding Jets: Modeling 3C 75}

\author{
S. M. Molnar\altaffilmark{1}, H.-Y. Schive\altaffilmark{2}, 
M. Birkinshaw\altaffilmark{3}, T. Chiueh\altaffilmark{4,5}
G. Musoke\altaffilmark{3}, and A. J. Young\altaffilmark{3}
}

\altaffiltext{1}{Institute of Astronomy and Astrophysics, Academia Sinica, P. O. Box 23-141,
                      Taipei 10617, Taiwan}

\altaffiltext{2}{National Center for Supercomputing Applications, Urbana, IL, 61801, USA}

\altaffiltext{3}{HH Wills Physics Laboratory, University of Bristol, Tyndall Avenue, Bristol BS8 1TL, UK}

\altaffiltext{4}{Department of Physics, National Taiwan University, Taipei 106, Taiwan}

\altaffiltext{5}{Center for Theoretical Sciences, National Taiwan University, Taipei 106, Taiwan}

\begin{abstract}
Radio observations suggest that 3C 75, located in the dumbbell shaped galaxy NGC 1128 
at the center of Abell 400, hosts two colliding jets. 
Motivated by this source, we perform three-dimensional hydrodynamical simulations 
using a modified version of the GPU-accelerated Adaptive-MEsh-Refinement 
hydrodynamical parallel code (\GAMER) to study colliding extragalactic jets.
We find that colliding jets can be cast into two categories: 
1) bouncing jets, in which case the jets bounce off each other keeping their identities, 
and 2) merging jets, when only one jet emerges from the collision.
Under some conditions the interaction causes the jets to break up into oscillating filaments of 
opposite helicity, with consequences for their downstream stability.
When one jet is significantly faster than the other and the impact parameter is small,
the jets merge; the faster jet takes over the slower one. 
In the case of merging jets, the oscillations of the filaments, in projection, 
may show a feature which resembles a double helix, similar to the radio image of 3C 75.
Thus we interpret the morphology of 3C 75 as a consequence of the collision of two jets with distinctly 
different speeds at a small impact parameter, with the faster jet breaking up into two oscillating filaments.
\end{abstract}


\section{Introduction}
\label{S:Intro}

Extragalactic jets are thought to play an important role in providing 
a heating mechanism in the center of cool core (CC) clusters of galaxies.
CC clusters exhibit a falling temperature and increasing gas 
density towards their center (for a review see \citealt{Fabian94}).
The high temperature (few million Kelvin) intracluster gas (ICG)
cools mainly via thermal bremsstrahlung and line emission, 
the cooling time being proportional to the inverse of the gas density.
The cooling times in the dense core of some CC clusters 
are much shorter than the Hubble time.

As a consequence of the short cooling times, we expect a large amount 
of cool gas and high star formation rate in these clusters.
However, X-ray observations made by \CHANDRA  and \XMM 
\citep[e.g.,][]{PetersonET2003,TamuraET2003,PetersonFabian2006}
did not find a large amount of cool gas in CC galaxy clusters,
and the observed star formation rate is much lower than predicted 
 \citep[e.g.,][]{Edge2001,SalomeCombes2003,McDonaldET2011,ODeaET2008}.
Non-gravitational energy input into CC clusters 
seems to be necessary to solve this, the so called ``cooling flow'', problem. 
This energy input is most likely self-regulated, since it should 
provide sufficient heat to quench the cooling, but at the same time 
it should not be so strong as to overheat and destroy the cool core. 
The most likely sources providing the sufficient heating 
to prevent the intracluster gas from runaway cooling 
are active galactic nuclei (AGNs) located at the center of CC clusters
\citep{McNamaraNulsen2007}, 
but the exact physical mechanism which distributes that heat is still not known.

Recent high resolution observations indicate that active galactic nuclei (AGNs) 
at the center of CC clusters of galaxies generate hot bubbles, jets, shocks, 
and turbulence in the ICG (see \citealt{Fabian2012ARAA} for a review).
Jets have been suggested as the main source of energy input from AGNs into the ICG. 
Numerical simulations have demonstrated 
that AGN feedback based on momentum-driven 
jets can prevent the cooling catastrophe 
\citep[e.g.,][]{GaspariET2011,MartizziET2012}.
Simulations assuming that the AGNs are powered by the accretion of cold gas 
produced stable thermal equilibrium and multiphase filamentary structures 
similar to those observed in nearby CC clusters 
\citep[e.g.,][]{GaspariET2012,LiBryan2014a,GaspariET2013}.
However, the cold gas forms an unrealistically massive stable disk in simulations,
suggesting that other physical processes are necessary to explain the 
observations \citep{LiBryan2014b}.

It was proposed and demonstrated by \cite{FalcetaET2010} that the combined 
effect of star formation and AGN feedback to prevent the cooling catastrophe.
More recently, \cite{LiET2015} demonstrated that 
momentum-driven AGN feedback and star formation
can prevent the cooling catastrophe 
producing self-regulated cycles of accretion by the central 
supermassive black hole, which heats the gas, 
followed by gas cooling until the next cycle. 
This model is mainly consistent with most observations
(but, e.g., occasionally, it does produce higher
star formation rate and cooling rate than observations imply).
However, more work is needed to improve the model parameters
and include more physics (e.g., transport processes and magnetic fields, 
which may provide additional pressure support, and likely suppress 
star formation and lower the star formation rate; \citealt{LooET2015}).
Magnetic fields are essential in launching and collimating the jets,
but perhaps less important for jets propagating on extragalactic scales
\citep[e.g.,][]{PudritzET2012}.
Most recent magneto-hydrodynamical simulations 
including AGN feedback, cooling, and anisotropic conduction
due to the jet magnetic field show that the main sources of
heating are still AGNs, however, conduction may contribute to heating 
significantly in the most massive clusters only if the maximum 
Spitzer conductivity is adopted along magnetic field lines 
\citep{YangReynolds2016ApJ818}.

\begin{figure}[t]
\includegraphics[width=.477\textwidth]{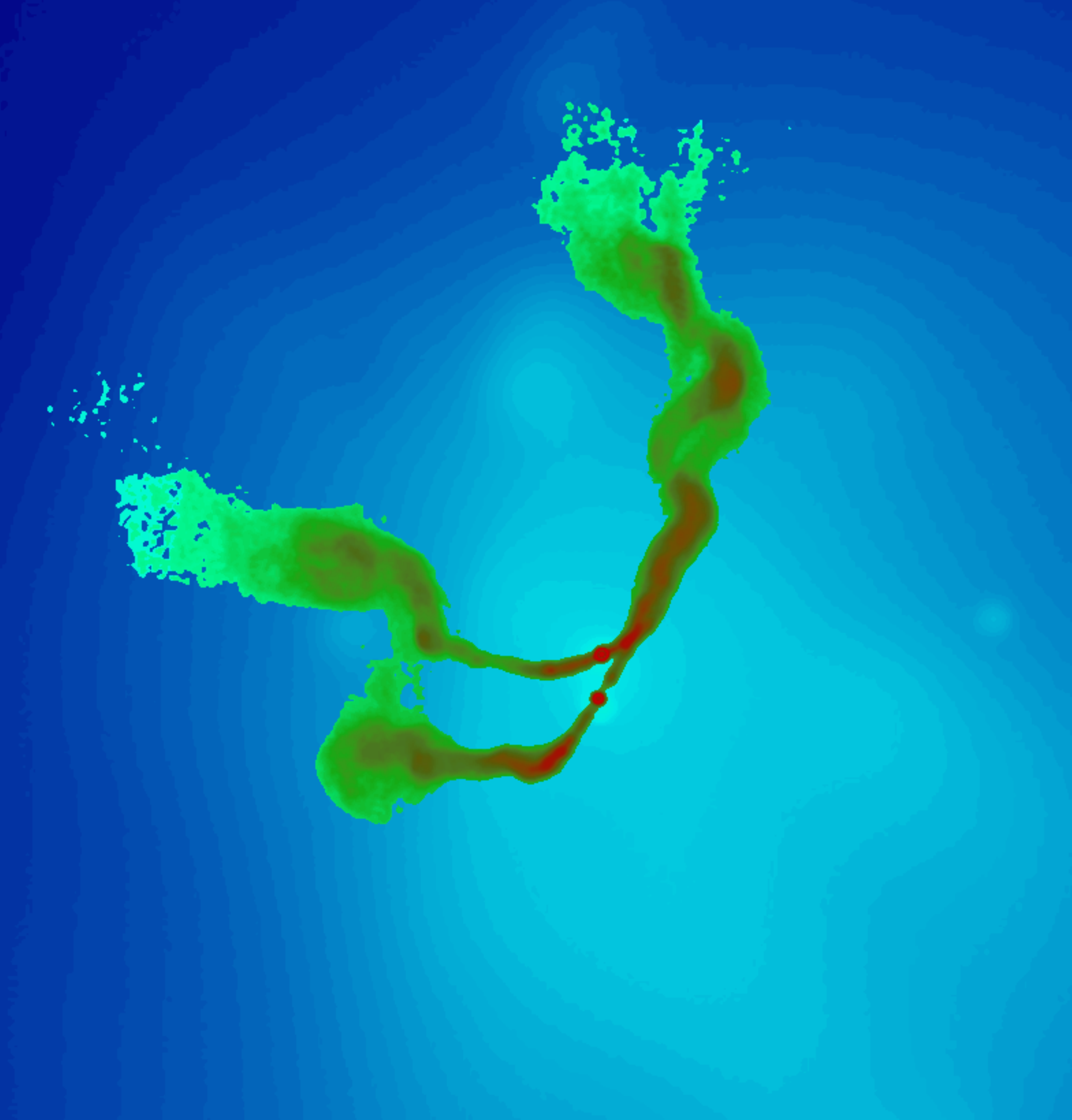}
\caption{
Superimposed radio (Very Large Array 8.4 GHz; green and red) and X-ray \CHANDRA in 0.7-7.0 keV; blue)
image of  \THREEC in NGC 1128 at the center of Abell 400 \citep[after][]{HudsonET06}.
The distance between the two AGNs is about 7 kpc in projection.
\vspace{0.3 cm}
\label{F:IMAGE3C75}
}
\end{figure}

The collisions of jets are extremely energetic events, thus they offer a unique opportunity to study 
astrophysical plasma under extreme conditions and infer the physical properties of jets based 
on the comparison of hydrodynamical (and magneto-hydrodynamical) simulations with observations. 
Jet collisions are also rare, but radio observations of the double twin-jet system, 3C 75, 
located in NGC 1128 at the center of the nearby galaxy cluster Abell 400, 
suggest that there are two colliding jets \citep[e.g.,][]{HudsonET06}.
In Figure~\ref{F:IMAGE3C75} we show superimposed radio and optical images of the centre of 3C 75.
The western jet from the northern AGN appears to collide with, and merge into, the northern jet from 
the southern AGN which does not seem to change its course significantly.

Motivated by 3C 75, we study the interaction between two bipolar jets by
performing hydrodynamical simulations using a modified version of the 
GPU-accelerated Adaptive-MEsh-Refinement (AMR)
hydrodynamical parallel code (\GAMER) developed at the Institute of Astrophysics of 
National Taiwan University \citep{SchiveET2010ApJS186}.

We present our results in the following sections. 
After this introduction we describe our numerical scheme (Section~\ref{S:JET_SIMULATIONS}).
In Section~\ref{S:RESULTS} we present the results of our hydrodynamical
simulations of colliding jets and provide a possible physical explanation 
for the twisted morphology of the jets systems of 3C 75. 
Section~\ref{S:CONCLUSION} contains our conclusions.

%
%

\section{Hydrodynamical Simulations of Colliding Jets}
\label{S:JET_SIMULATIONS}

The formation of jets around AGNs and their propagation out 
of the host galaxy into the intracluster environment involves complicated 
non-linear physical processes including non-gravitational physics on a 
large dynamical range in the space and time domain. 
Presently it is not feasible to carry out numerical simulations to cover
all relevant physical process from the formation and propagation of extragalactic jets
from a scale of one pc to tens of kpc.
We expect that the jets are launched at relativistic speeds ($\sim 0.98c$)
from their AGNs, but they are significantly decelerated by the time they reach
the edge of their host galaxies ($\sim 0.4c$), and further decelerate in the intergalactic medium
\citep[e.g.,][]{LaingBridle2014MNRAS}. 
In our study we focus on simulating collisions between jets from FR I sources.
FR I jets are expected to decelerate significantly, to v$_{jet} \simless\, 0.1c$
on kpc scales \citep[e.g.,][]{FerettiET1999,LaingET1999}.
We modeled the jets as bipolar outflows of high temperature, light fluid launched from cylindrical 
nozzles at the center of our simulation box with velocities that are expected for these jets when 
they propagate to extragalactic distances.
Since we deal with non-relativistic jet velocities, 
we may adopt the equations of ideal non-relativistic hydrodynamics.

In our first study of jet collisions, we assumed an initially homogeneous isothermal 
ambient gas and a continuous inflow of gas with constant injection velocity, and 
fixed density and temperature for each of the two stationary jets.

We used a modified version of the GPU-accelerated Adaptive-MEsh-Refinement (AMR)
hydrodynamical parallel code (\GAMER) developed at the National Taiwan University 
\citep{SchiveET2010ApJS186,Schive2012} to carry out our simulations of colliding jets. 
\GAMER adopts a novel approach in improving the performance of astrophysical numerical simulations.
It uses GPUs for solving the partial differential equations, for example, hydrodynamics,
self-gravity, and magnetohydrodynamics (Zhang et al., in preparation), and uses CPUs
for manipulating the AMR data structure. The parallel performance has been highly
optimized by implementing 
(1) hybrid MPI/OpenMP/GPU parallelization, 
(2) concurrent execution between multiple CPUs and GPUs, 
(3) asynchronous data transfer between CPUs and GPUs, and 
(4) Hilbert space-filling curve for load balancing. 
The code supports a variety of hydrodynamic schemes. In this work we adopt
the corner-transport-upwind method \citep{Colella1990}, Roe's Riemann solver \citep{Roe1981},
and the piecewise parabolic data reconstruction \citep{CW1984}.
The overall performance using these schemes is measured to be $\sim 7\times10^7$
cell updates per second on a single NVIDIA K40 GPU, which is found to be two orders
of magnitude faster than a single CPU core and one order of magnitude faster
than a ten-core CPU using an Intel Xeon E5-2670 v2 CPU at 2.50 GHz.
The accuracy of \GAMER has been carefully verified \citep{SchiveET2010ApJS186,Schive2012}.
We also tested \GAMER by comparing results of bipolar single jet simulations carried out 
with \GAMER and the publicly available and extensively tested Eulerian parallel code \FLASH
developed at the University of Chicago \citep{FryxellET2000}.

\GAMER solves the Euler equations in conservation form. 
In this case, in Cartesian coordinates, $(x_1,x_2,x_3)$,
the mass, momentum and energy conservation equations may be expressed as:
\begin{eqnarray}   
  \frac{\partial\rho}{\partial t}+\frac{\partial}{\partial x_j}(\rho \bv_j)  & = &  0              \label{E:EULER1} \\     
  \frac{\partial(\rho \bv_i)}{\partial t}+\frac{\partial}{\partial x_j} 
                                                        (\rho \bv_i \bv_j+p \delta_{ij})   & = & 0 \label{E:EULER2} \\
  \frac{\partial e}{\partial t}+\frac{\partial}{\partial x_j}[(e+p)\bv_j]    & = & 0               \label{E:EULER3} 
,
\end{eqnarray}
where $\rho$, $p$, and $e$ are the mass density, the thermal pressure,
and the total energy density of the gas, $e = (1/2) \rho \bv^2+\epsilon$, where 
$\epsilon$ is the internal energy density,
$\bv_i$ and $\bv_j$ are the components of the flow velocity vector, ${\bf v}$, and
the spatial indices, $i$, and $j$ take a value of 1, 2, or 3. 
All dependent physical variables are evaluated at position $ {\bf r} = (x_1,x_2,x_3)$ 
and time $t$ (we suppressed the independent variables for clarity).
This set of equations is closed by using the equation of state for ideal gas:
\begin{equation}
   p = (\gamma-1) \, \epsilon
,
\label{eq:EOS}
\end{equation}
where $\gamma$ is the ratio of specific heats.
For monoatomic classical and ultra-relativistic ideal gases, 
the adiabatic index changes between 5/3 and 4/3.
We adopt $\gamma \le 5/3$ as a good approximation, 
since we assume that the jet temperatures are in the non-relativistic regime 
($T_{jet} \ll 10^5$ keV; e.g., \citealt{MignoneMcKinney2007MNRAS}).

\subsection{Initial Conditions for Colliding Jets}
\label{SS:InitialConditions}

We preformed simulations of two bipolar jets with different injection velocities, cross sections, 
densities, and temperatures for the jets and the ambient gas.
In this paper we present simulations relevant to our qualitative analysis of colliding
extragalactic jets.

With future applications in mind, we chose the physical parameters of our jet 
system to match those derived for 3C 75 based on radio and X-ray observations \citep{HudsonET06}.
We fixed the initial density and temperature for the ambient 
gas at values suggested by the observations of 3C~75, and adopted a gas density
and temperature of $\rho_{amb} = 5 \times 10^{-29}\,$g\,cm$^{-3}$ and $T_{amb} = 0.15\,$keV.
We simulated both sides of the two bipolar jets to model 3C~75 in a 
simulation box size of 20 kpc $\times$ 100 kpc $\times$ 20 kpc, 
($[x_1,x_2,x_3] = [x,y,z] = [\pm 10, \pm 50,\pm 10]$ kpc)
choosing the $y$ axis as the direction of the main jet propagation. 
We used outflow boundary conditions which are suitable for colliding jets.

\begin{figure}[t]
\centering
  \includegraphics[width=0.25\textwidth]{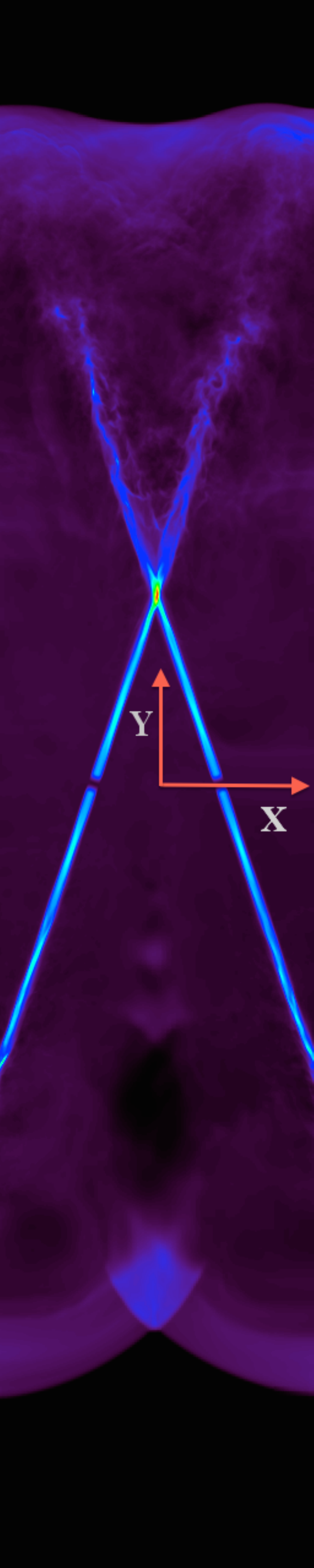}
\caption{ \protect
 Total energy density of our colliding jet model with velocities of v$_1$= v$_2 = 18000\,$\KMSEC
 and an impact parameter of P = 0.6 in units of the jet radius (see Section~\ref{SS:InitialConditions};
 model P060V18) projected to the $(x,y)$ plane.
 The simulation box size is $(x,y,z) = (\pm 10, \pm 50,\pm 10)$ kpc.
\label{F:TEST47ENER}
}
\end{figure}

The jets were continuously ejected from two nozzles aligned with the $(x, y)$ plane 
centered at $y=0$ with a fixed distance of 8$\,$kpc between them (in $x$) 
and an offset in the $z$ direction: 
${\bf r_{1.2}} = (\pm 4\,\rm{kpc},0,\pm z_0)$, 
where $z_0$ defines the impact parameter, ${\rm P} = 2 z_0$, the offset between the 
two centers of the nozzles. 
We illustrate our computational setup in Figure~\ref{F:TEST47ENER}, where we show 
the total energy density of one of our models (model P060V18; see Table~\ref{T:TABLE1}) 
projected to the $(x,y)$ plane. The $z$ axis is pointing out of the $(x,y)$ plane with
$(x,y,z)$ forming a right handed Cartesian coordinate system.
Hereafter we will express the impact parameter in units of the jet diameter, ${\rm P}= 2 z_0/ 2 r_{jet}$,
thus P = 0 means a head-on collision. When the two jets do not encounter, P $\ge 1$.
With a viewing angle of 20\DEG, this gives a projected distance of 7.5 kpc  
for the jet sources (AGNs), which is in agreement with that of 3C 75 \citep{HudsonET06}.

We adopt a setup for initial conditions suggested by the observed morphology of 3C 75.
All intrinsic parameters describing the injection of the two jets were the same: 
the jet radius: $r_{jet}= 0.35\,$kpc, the length of the nozzle: $0.7\,$kpc
(the radius and the height of the cylindrical nozzles).
The directions of the jet injection velocity vectors were fixed in the $(x, y)$ plane
with directions: ${\bf n}_{1,2} = (\pm1, 3, 0)/\sqrt{10}$.
The velocity field within each nozzle generating the bipolar jets was changed smoothly 
from -v$_i$ to +v$_i$ ($i = 1,2$) towards the negative and the positive $y$ axis 
to avoid a large unphysical jump between velocities within adjacent simulation cells,
which could cause numerical problems.
We adopted $\rho_{jet} = 10^{-30}\,$g\,cm$^{-3}$ and $T_{jet} = 12.2\,$keV for the 
gas of the jets, so the jet is roughly in pressure equilibrium with the ambient medium.

We ran jet collision simulations with a set of different impact parameters changing $z_0$,
and, fixing one jet velocity at v$_1 = 18000\,$\KMSEC, with different velocities of the 
second jet from v$_2 = 10000$ \KMSEC to v$_2 = 18000\,$\KMSEC
(internal jet Mach numbers, M$_{\rm jet} =\,$v$_2/c_{s2}$, where $c_{s2}$ is the sound speed 
inside the second jet, 4.4 $\le {\rm M}_{\rm jet} \le 7.9$; 
see the input parameters and the IDs for our different models in Table~\ref{T:TABLE1}).
In order to study the jet interactions considering an extensive range of values of the impact parameter,
we also ran an additional simulation with a large impact parameter (P = 1.3; model P130V18).
We ran most of our hydrodynamical simulations for 50 Myr. In those cases when 
we adopted slow injection velocities for the second jet (v$_2 = 10000, 12000\,$\KMSEC),
we ran the simulations for 70 Myr in order to make sure that the slower jet propagates 
well away from the collision region.

\section{Results}
\label{S:RESULTS}

\subsection{Bouncing and Merging of Extragalactic Jets}
\label{SS:MERGING}

We show the results of our colliding jet simulations in 
Figures~\ref{F:PROJZENERALL} and \ref{F:PROJYENERALL}
as projections of the total energy density on the $(x,y)$ plane, the plane of jet propagation, 
and the $(x,z)$ plane, showing the interacting jet offsets ($\pm \,z_0)$.
The columns of these figures show simulations wih different impact parameters, 
${\rm P} = 0$, 0.15, 0.3, 0.45, 0.6, and 0.8 times the diameter of the jet (left to right). 
The rows show the simulations at different velocities for the second jet expressed
as the jet Mach number, M$_{\rm jet} = 7.9$, 7.0, 6.1, 5.3, and 4.4 (top to bottom;
see Section~\ref{SS:InitialConditions} and Table~\ref{T:TABLE1} for details).

The collision regions can be seen clearly in these figures as an enhancement in the 
projected energy marked by red. 
In each panel in Figure~\ref{F:PROJZENERALL}, 
the two jets are moving from left to right nearly parallel to the horizontal axis (simulation $y$ axis).
In Figure~\ref{F:PROJYENERALL} the first jet with fixed Mach number  (M$_{\rm jet} = 7.9$)
is moving left to right, the second jet, with different velocities, is moving in the opposite direction.

In the first column in Figures~\ref{F:PROJZENERALL} and \ref{F:PROJYENERALL}
we show jet collisions with zero impact parameter (P = 0). 
In all these cases only one jet seems to emerge after the collision (see Figure~\ref{F:PROJZENERALL}). 
However, because of symmetry, after the collision these jets are somewhat 
spread out in the out-of-collision-plane direction ($z$; see Figure~\ref{F:PROJYENERALL}).

In most of those cases where we adopted a finite impact parameter (P $> 0$),
two jets can be seen after the collision.
As a consequence of the large velocities, the jets bounce 
off each other and survive as individuals (see Figure~\ref{F:PROJZENERALL}). 
In Figure~\ref{F:PROJZENERALL} we can see that after the collision, the jets show
increasing instabilities as the impact parameter is decreased, and 
in some cases, as a result of growing instabilities, the jets break up into oscillating filaments.

After the collision the two jets still propagate in roughly the same plane
(see Figure~\ref{F:PROJYENERALL}), but the plane is rotated relative to the horizontal direction 
by an angle which is larger for smaller impact parameters and larger velocity differences 
between the two jets.

In those cases when the second jet is much slower than the first, 
the two jets may merge, and the bending angle after collision of the slower jet is larger 
for higher velocity differences between the two jets (see Figure~\ref{F:PROJZENERALL}).

Based on our results, we can classify colliding jets into two
categories: (1) bouncing and (2) merging jets.
We illustrate the difference between these two categories in 
the first and second columns in Figure~\ref{F:TEST3356PROJZ}.
In this figure we display 
the total energy density, 
scaled entropy per particle ($K \propto T/\rho^{2/3}$), 
enstrophy, $\Omega  = (1/2)\, \omega^2$ (see Section~\ref{SS:ENSTROPHY}),
and helicity, ${\bf v} \cdot \bomega$, 
where $\bomega$ is the vorticity, $\bomega  = \nabla \times {\bf v}$,
projected to the $(x, y)$ plane (from top to bottom, first two columns).
The first column illustrates bouncing of two fast jets with v$_1= $ v$_2 = 18000\,$\KMSEC,
and a large impact parameter, P = 0.8 (model P080V18);
the second column shows projections of a merging jet simulation with a small impact parameter,
P = 0.3, and large velocity difference: v$_1 =18000\,$\KMSEC vs. v$_2 = 10000\,$\KMSEC (model P030V10).
The jets propagate from left to right. 
The point of collision is in the middle of the left hand side of the energy panels marked with red
(see top left panel for each simulation).
It can be seen clearly from Figure~\ref{F:TEST3356PROJZ} that, after the collision,
the bouncing jets (1st column) keep their identities and travel relatively unaffected, 
while, in the case of merging jets, after the collision, the faster jet absorbs the slower jet, 
although not instantly (2nd column).

\begin{figure}[t]
\includegraphics[width=.48\textwidth]{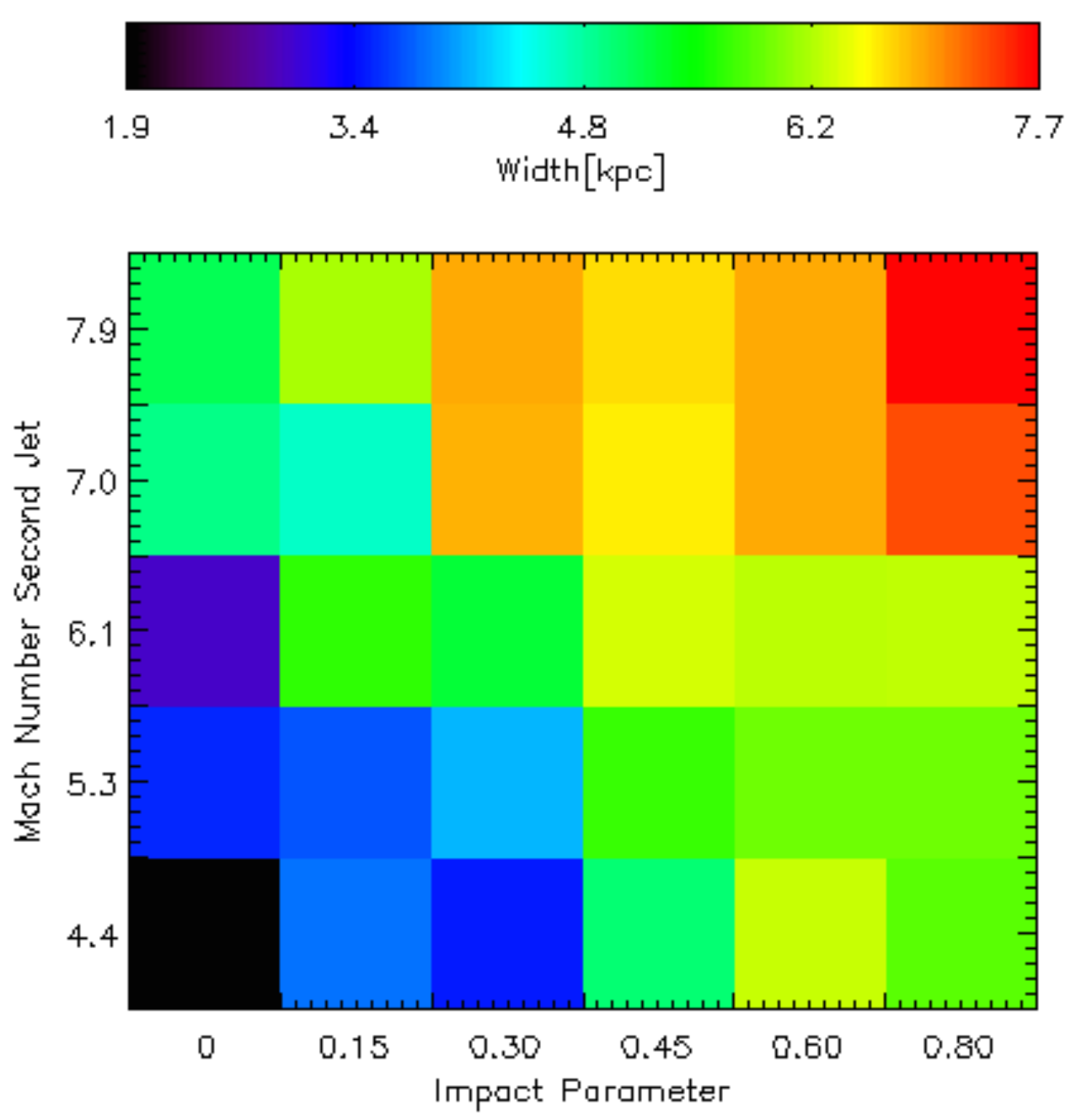}
\caption{
The spread around the maximum value of the total energy of colliding jets 
projected to the $(x, y)$ plane, $e_{z}(x,y)$.
The color scale represents the width of $e_{z}$ 
where $e_{z} \ge 5$\% of its maximum value at a distance of 15 kpc from the collision point 
($e_{z}[15,y] \ge 0.05\, \times {\rm Max}\{e_{z}[15,y]\}$; see Section~\ref{SS:MERGING})
as a function of impact parameter (P = 0, 0.15, 0.3, 0.45, 0.6 and 0.8 in units of the diameter of the jet)
and the Mach number of the second jet (M$_{\rm jet}$ = 4.4, 5.3, 6.1, 7.0, and 7.9).
\label{F:QPHASEDIAGRAM}
}
\end{figure}

We quantify the effect of jet collisions by projecting the total energy density of the system 
to the $(x, y)$ plane, $e_{z}$, and measuring the spread 
of $e_{z}$ along a line parallel to the $x$ axis 
(i.e, along a vertical line in Figure~\ref{F:PROJZENERALL})
at a distance of 15 kpc from the collision point ($e_{z}[15,y]$),
and identify the two regimes of the parameter space resulting in 
bouncing and merging jets. 
In Figure~\ref{F:QPHASEDIAGRAM} we plot the width of the projected total energy density
where it is $\ge$ 5\% of its maximum value ($e_{z}[15,y] \ge 0.05 \times {\rm Max}\{e_{z}[15,y]\}$)
in a grid of impact parameter (0 $\le$ P $\le 0.8$) and the velocity of the second jet 
(expressed as the Mach number, 4.4 $\le {\rm M}_{\rm jet} \le 7.9$).
The black-blue-green-orange-red colors represent increasing spread of the jet.
When the jets merge the spread is limited because the second jet disappears
(black and dark blue color pixels).

A trend can be seen clearly in Figure~\ref{F:QPHASEDIAGRAM}: 
the spread increases as the impact parameter and 
the velocity of the second jet increases. 
The jets merge when the impact parameter and the velocity of the second jet 
are small (magenta and dark blue pixels; P000V10, P000V12, P000V14, P015V10, P015V12, 
and P030V10, see Table~\ref{T:TABLE1}), otherwise they bounce. 
Since the $x$ components of the velocities of the jets have opposite signs,
the relative value of this component decreases (it should be zero for v$_1$ = v$_2$),
and, as a consequence, the angle between the bouncing jets decreases as does 
the spread between them.
This is the reason for the increasing spread with increasing impact
parameter (left to right in Figure~\ref{F:QPHASEDIAGRAM}).
Even though both jets propagate in the positive $y$ direction,
the $y$ component of the velocities of both jets decreases because, 
as they collide, they generate more turbulence and drive more waves 
as they propagate due to the increased instabilities.

\subsection{Instabilities in Colliding Jets}
\label{SS:Instabilities}

In general, as the jets propagate in the ambient medium,
instabilities are generated due to the Kelvin-Helmholtz process
\citep[e.g.,][and references therein]{Birkinshaw1991,Ferrari1998}.
The form of the instabilities depends on the thickness
of the contact layer between the jet and the ambient medium.
Near the origin of the jet the transition zone is thin,
and the vortex sheet approximation can be used assuming
large density and velocity gradients. 
Farther from the origin, the contact layer extends due
to matter entrainment, turbulence and other non-linear effects.

There are two main types of instability in jets:
the surface modes, which have steeply decreasing amplitude
as a function of the distance from the jet surface, and
reflected body modes, which affect the entire gas in the jet.
Reflection modes dominate if M$_{\rm jet} \, \simgreat \,2 \, {\sqrt 2}$
\citep[e.g.,][]{Ferrari1998}.
The jet Mach numbers in our hydrodynamical simulations 
range from 4.4 to 7.9, therefore we expect the reflection modes to dominate.
The wavelengths of the reflection modes can be estimated as 
$\lambda \sim 2 \pi R \, {\rm M}_{\rm jet}$, 
where $R$ is the characteristic cross section of the jet.
In our case, $\lambda \sim 17\,$ kpc. 
Our simulations show long wavelength oscillations along the jets 
with wavelengths in the range of 10 - 20 kpc, which is compatible to the expected wavelength
of the reflection modes.

\begin{figure*}[t]
\centering
  \includegraphics[width=0.326\textwidth]{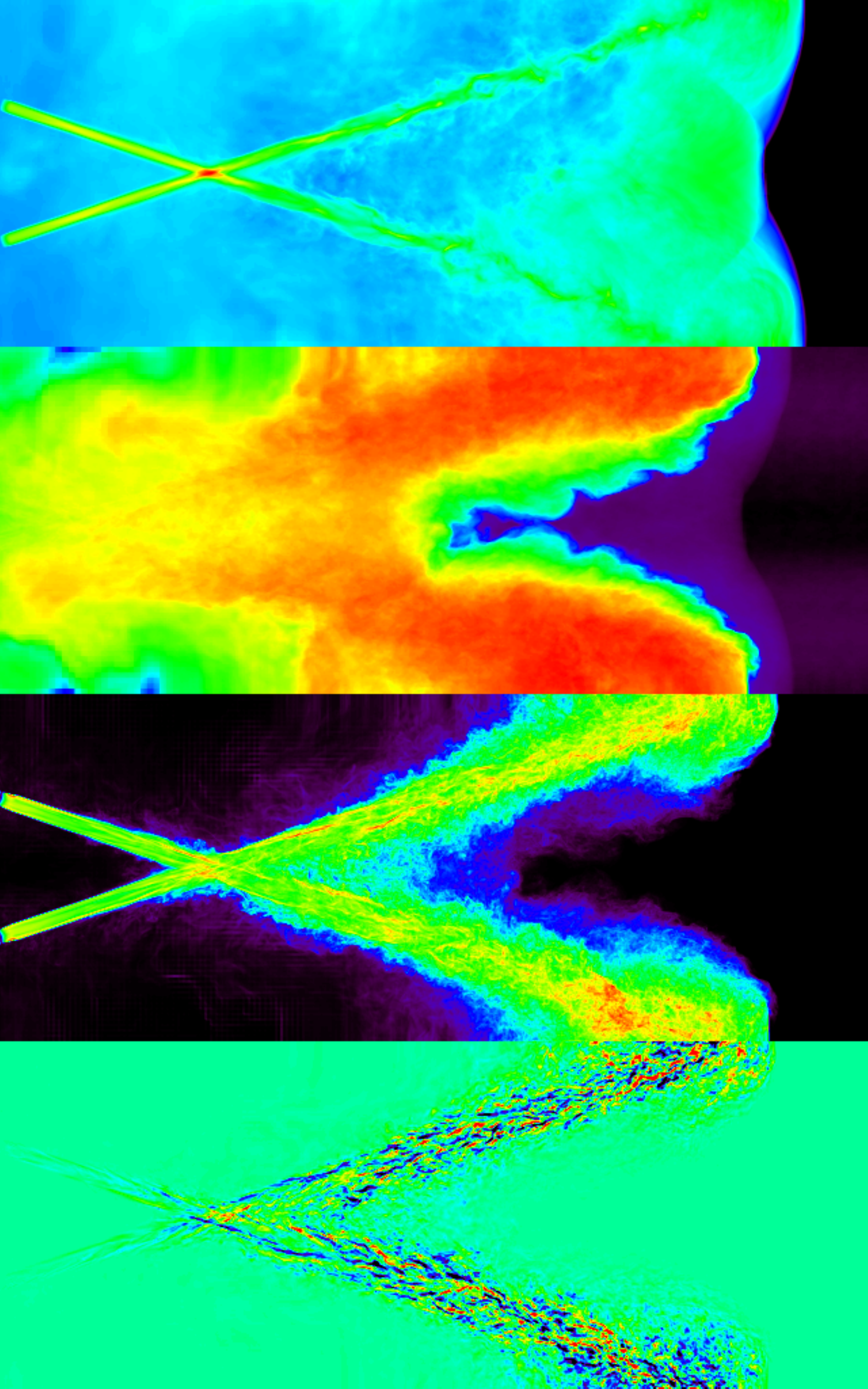}  
  \includegraphics[width=0.326\textwidth]{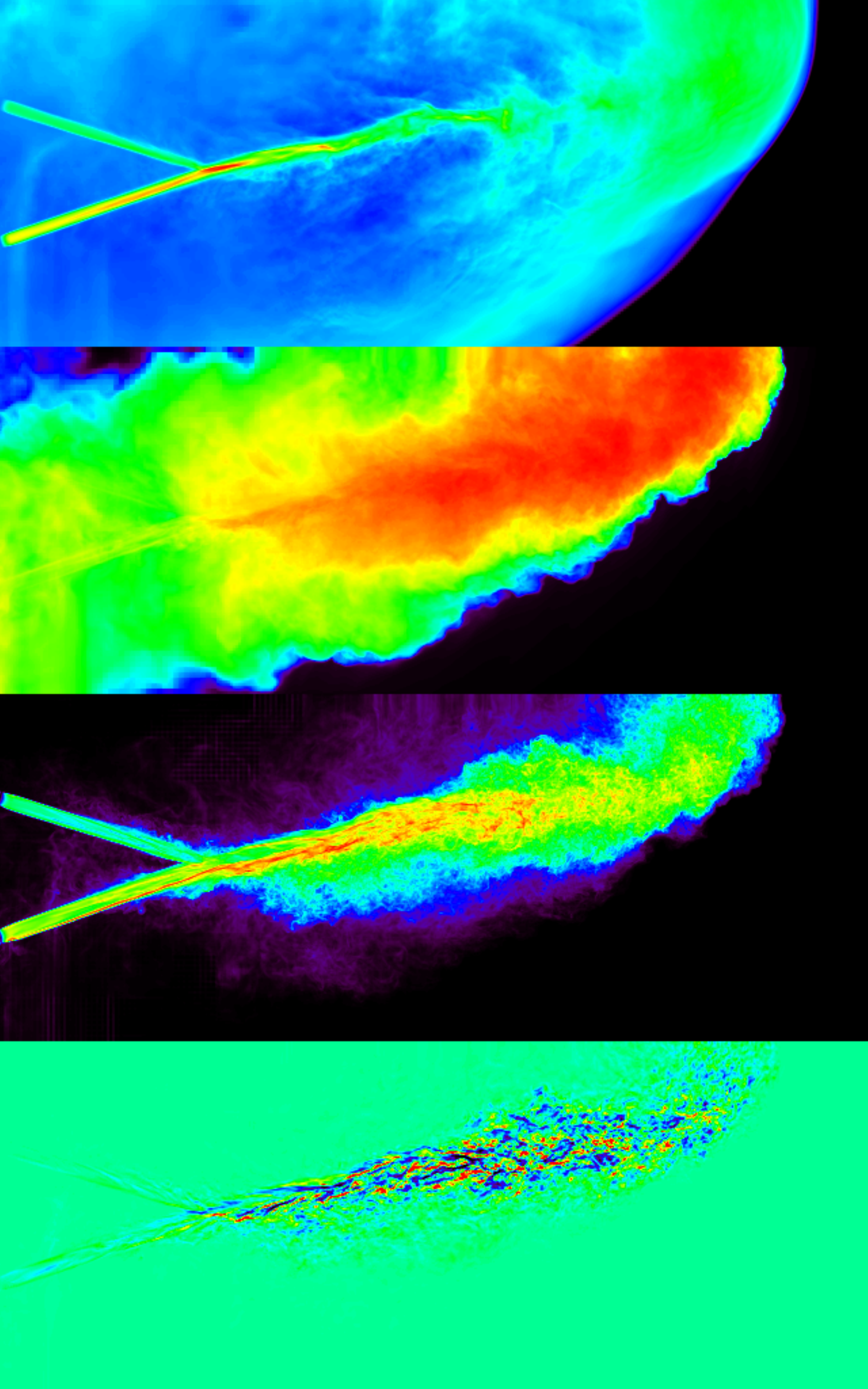}  
  \includegraphics[width=0.33  \textwidth]{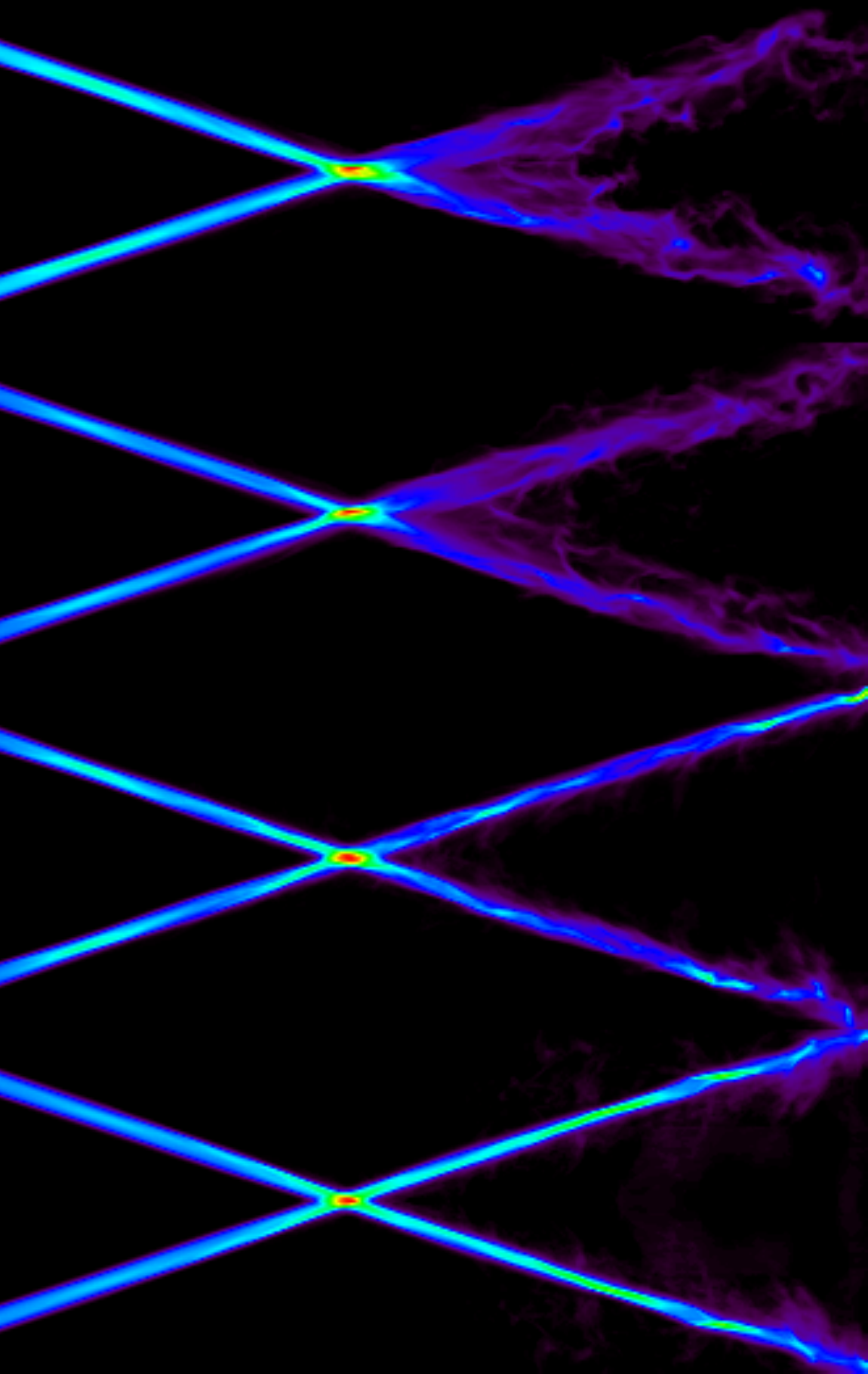}
\caption{ \protect
 Physical parameters of bouncing and merging jets projected to the $(x, y)$ plane.
 The jets propagate from left to right. 
 The 1st column shows 
 the total energy density, scaled entropy per particle, enstrophy (all in log scale), and helicity 
 (linear scale) from our bouncing of jet simulation with v$_1$= v$_2 = 18000$\KMSEC,
 and a large impact parameter, P = 0.8. The 2nd column displays projections of the same physical quantities 
 from our merging jet simulation with a small impact parameter,
 P = 0.3, and large velocity difference: v$_1 =18000\,$\KMSEC, v$_2 = 10 000\,$\KMSEC.
 Panels in the 3rd column show the total energy projected to the $(x, y)$ plane 
 from our bouncing jet simulations with fixed jet velocities of v$_1$ = v$_2$ = 18000 \KMSEC,
 but different impact parameters: 
 P = 0.5, 1.0, 1.3, and $\infty$ in units of the jet diameter (see Section~\ref{SS:InitialConditions}).
\vspace{0.3 cm}
\label{F:TEST3356PROJZ}
}
\end{figure*}

In the 3rd column in Figure~\ref{F:TEST3356PROJZ} we show the total energy projected to the 
$(x, y)$ plane from our colliding jet simulations with fixed jet velocities: 
v$_1$ = v$_2 = 18000\,$\KMSEC, but different impact parameters: 
P = 0.5, 1, 1.3, and with P = $\infty$ representing two non-interacting pairs of jets 
(top to bottom; models: P045V18, P100V18, P130V18, and SNGV18).
In the case of non-interacting jets (see bottom panel in the 3rd column),
linear instabilities grow non-linear only after the jets travel about 20 kpc.
However, it can be seen from the 3rd column in Figure~\ref{F:TEST3356PROJZ} 
that the instabilities are enhanced in 
colliding jets and they grow non-linear soon or immediately after the collision  
(see 1st, 2nd and 3rd panels).
As we increase the impact parameter, the instabilities decrease (top to bottom in the 3rd column; 
see also the trend from left to right in the different columns in Figure~\ref{F:PROJZENERALL}).

The jets in Figure~\ref{F:TEST3356PROJZ} 
do not seem to expand where they meet, therefore the jets in
our simulations with P = 1 and 1.3 do not actually collide.
However, interestingly, comparing the 2nd and 3rd panels (P = 1 and 1.3) 
in Figure~\ref{F:TEST3356PROJZ}
with the last panel (P = $\infty$), we notice that the two jets still interact with each other 
via turbulence, which enhances the instabilities and generates oscillations 
traveling down the jets.

Studying the images from our hydrodynamic simulations, 
we notice that in some cases the jets break into filaments due to instabilities 
(e.g., mo dels P080V18, P100V18, and SNGV18, columns 4, 5, and 6 top panels,
and models P100V10, and SNGV10, columns 5 and 6 bottom panels
in Figure~\ref{F:PROJZENERALL}).
Our simulations show that, in general, filaments form in jets only far
from the source ($\sim$30 kpc) unless the collision is strong,
i.e, the impact parameter is small (e.g., models P045V18 and P080V18; 
compare top panels in columns 4 and 6).
However, filaments of this form are not frequently seen in radio images of extragalactic jets,
which is most likely due to the fact that the jets usually reach lower-pressure regions 
in the ambient medium and expand substantially before filaments could develop.

\subsection{Enstrophy Generation in Colliding Jets}
\label{SS:ENSTROPHY}

\begin{figure}[t]
\includegraphics[width=.45\textwidth]{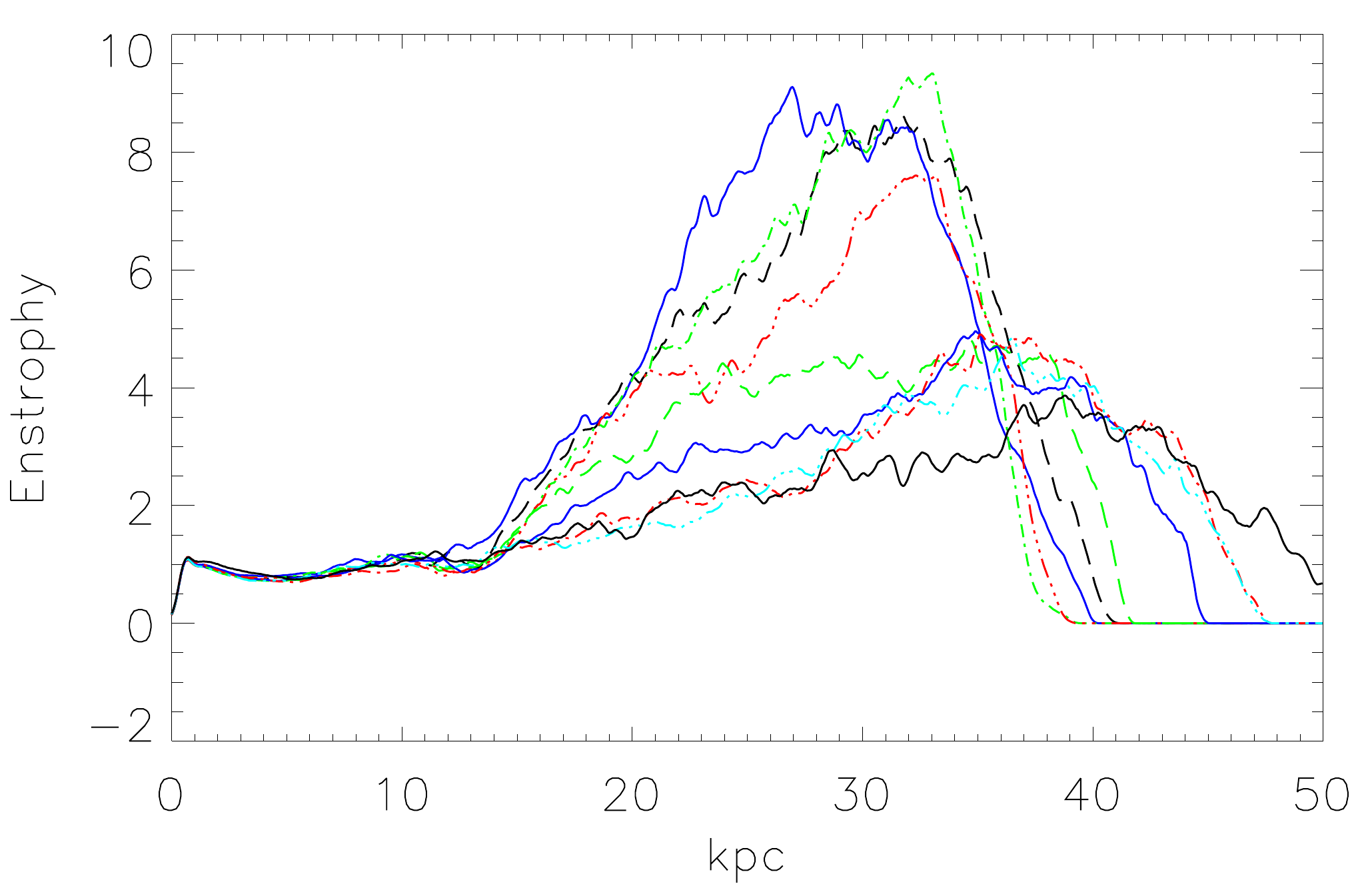}
\includegraphics[width=.45\textwidth]{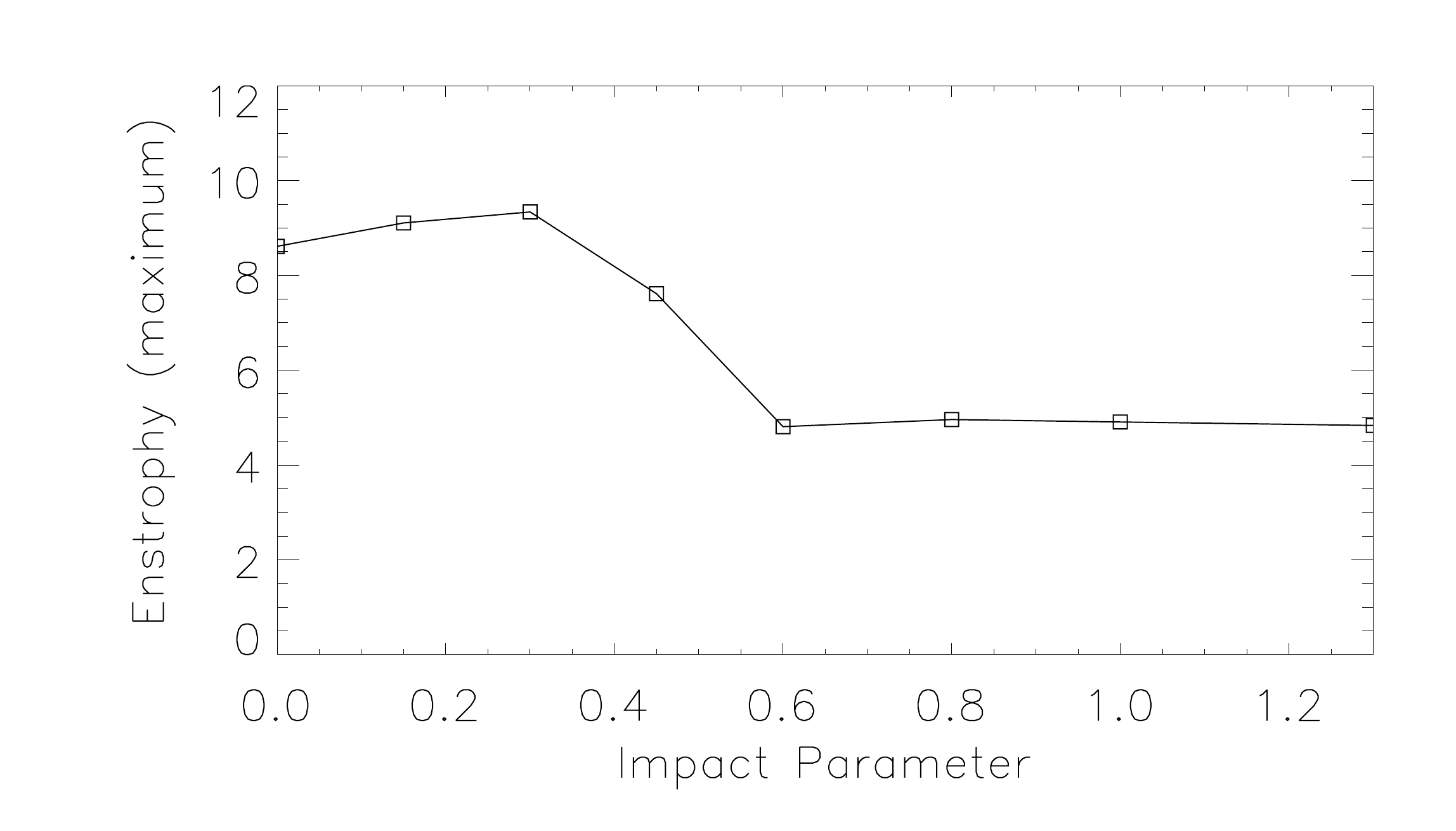}
\caption{
      Top Panel: Projected enstrophy as a function of distance 
      along the jet from our hydrodynamical simulations of colliding jets 
      with different impact parameters: P = 0, 0.15, 0.30, 0.45 
      (black/dashed, blue/solid, green/dash-dotted, red/dash-dot-dot-dotted lines)
      with enstrophy maximum greater than 6, and P = 0.60, 0.80, 1.00, and 1.3 
      (green/dashed, blue/solid, red/dash-dot-dot-dotted, and aqua/dash-dotted lines) with maximum less than 6
      in units of the diameter of the jet with fixed injection velocities: v$_1$ = v$_2$ = 18000 \KMSEC.
      The black/solid line extending farther than 50 kpc represents non-interacting jets (P = $\infty$).
      The enstrophy is normalized to its value at injection for each model. 
      Bottom Panel: The maximum values of the enstrophy as a function of the impact parameter 
      (from data displayed in the top panel).
\label{F:ENSTROPHYV18}
}
\end{figure}

The effect of interaction between jets due to turbulence generated by collision
can be quantified using the enstrophy density, the mean squared vorticity, 
\begin{equation}   \label{E:ENSTROPHY}
   \Omega ({\bf r}) =  \frac{1}{2} \bigl| \bomega ({\bf r})  \bigr|^2 
,
\end{equation}
where  $\omega ({\bf r}) = \nabla \times {\bf v}({\bf r})$ is the 
vorticity at position ${\bf r}$ (note that, occasionally, it is defined without the 1/2 factor).
In Figure~\ref{F:ENSTROPHYV18} we show the integrated enstrophy (over $x$ and $z$)
as a function of distance from the jet source (the $y$ coordinate) for different impact parameters:
P = 0, 0.15, 0.30, 0.45, 0.60, 0.80, 1.00, 1.3, and $\infty$ in units of the jet diameter.
The maximum distance to which the enstrophy shows the jets to propagate as coherent structures 
increases monotonically as the impact parameter is increased, while the enstrophy peak drops
significantly at ${\rm P} > 0.5$, as the jet centers become separated by more than one jet radius.

As the jets propagate, most of the enstrophy (turbulence) is generated
near the jet (3rd panel in the first and second columns in Figure~\ref{F:TEST3356PROJZ}).
In the top panel in Figure~\ref{F:ENSTROPHYV18} we show the projected enstrophy, 
normalized to its value at injection for each model, 
as a function of the distance along the jet for $y \ge 0$ from the simulations 
of colliding jets with fixed jet velocities, v$_1$ = v$_2$ = 18000 \KMSEC,
but with different impact parameters: $0 \le {\rm P} \le \infty$ (see last row in Table~\ref{T:TABLE1},
and P130V18). 
It can be seen from this figure that more enstrophy is generated right after the 
collision when we adopt smaller impact parameters. The largest enstrophy peaks ($> 6$) belong to 
the four smallest impact parameters (P = 0, 0.15, 0.30, 0.45).

The bottom panel of Figure~\ref{F:ENSTROPHYV18} shows the maximum of the enstrophy as a 
function of the impact parameter for interacting jets ($0 \le {\rm P} \le 1.3$).
The enstrophy maximum shows a ``phase transition'' as a function of the impact parameter
at a critical point, ${\rm P_{crit}} \sim 0.5$.
Small impact parameters , $0 \le {\rm P} \lesssim {\rm P_{crit}}$, result in large enstrophy maxima, 
$\sim 9$, in the transition  region around ${\rm P_{crit}}$, the enstrophy drops to about half of this value, 
and for large impact parameters, ${\rm P_{crit}} \lesssim {\rm P} \le 1.3$, the maximum is $\sim 4.6$.
A likely explanation for this ``phase transition'' is the following.
At the collision region of the two jets, the jets slightly oscillate in direction
as a result of the growing Kelvin-Helmholtz instabilities.
These oscillations induce a change in the jet gas mass involved directly 
in the collision by changing the collision cross section.
The larger the changes in the mass involved in the collision, 
the more turbulence is generated driving more shock waves into the 
ambient medium resulting larger enstrophy maxima.
The change in the mass involved in the collision is larger for small 
impact parameters, generating more turbulence.
As we increase the impact parameter, the change in the jet gas mass 
due to the oscillations of the jets drops at around P $\sim 0.5$, and, as a consequence,
the amount of turbulence and thus the enstrophy maximum are also
reduced significantly.

As the jets are plunging through the ambient gas, their bulk kinetic energy
dissipates into shocks and turbulence, and, as a result, they slow down.
The dissipation of the bulk kinetic energy into turbulent energy
can be quantified using the enstrophy density.
The scalar product of the curl of Equation~\ref{E:EULER2} with $\bomega$ leads to
the equation for enstrophy generation along the jet,
\begin{equation}   \label{E:ENSTRGEN}
  \frac{D \Omega }{D t} = \bomega \cdot \bigl( \bomega \cdot \nabla \bigr) {\bf v} \,
                                       -  2 \Omega  \bigl( \nabla \cdot {\bf v} \bigr)
                                       +\, \bomega \cdot \bigl(\nabla \rho \times  \nabla p \bigr)/ \rho^2  
,
\end{equation}
where the left hand side of this equation is the Lagrangian derivative, 
$D\, \Omega / D\, t = \partial\,  \Omega / \partial\, t +  ({\bf v} \cdot  \nabla)\,\Omega$, and 
the source terms on the right hand side are the enstrophy stretching term, the compression term, 
and the baroclinic contribution, which is a result of the misalignment between
the gradients of the gas density and the pressure 
\citep[e.g.,][]{PorterET2015}.

\begin{figure}[t]
\includegraphics[width=.45\textwidth]{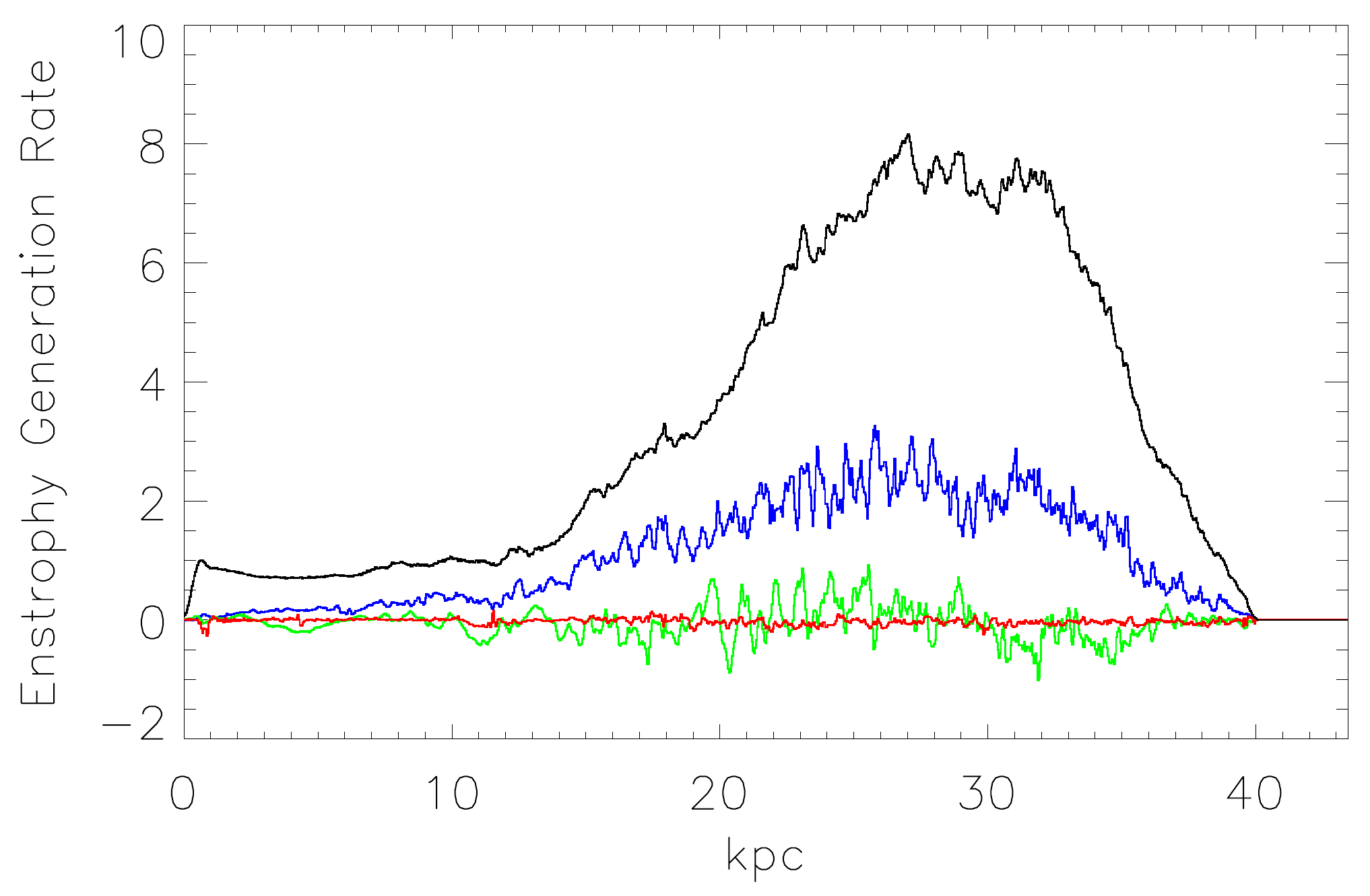}
\caption{
      Enstrophy generation rate as a function of distance along the jet from our hydrodynamical 
      simulations of colliding jets with impact parameter P = 0.15 and 
      injection velocities v$_1$ = v$_2$ = 18000 \KMSEC.
      The blue, green and red lines represent the relative contributions to the 
      enstrophy generation from stretching, compression, and from the baroclinic term 
      (the 1st 2nd and 3rd terms in Equation~\ref{E:ENSTRGEN}).
      For comparison, we also show the enstrophy (black solid line) normalized to its value at the injection
      (also shown in Figure~\ref{F:ENSTROPHYV18}). 
\label{F:ENSTROPHGEN}
}
\end{figure}

We illustrate the generation of enstrophy using our simulation with a small 
impact parameter, P = 0.15, and injection velocities v$_1$ = v$_2$ = 18000 \KMSEC.
In Figure~\ref{F:ENSTROPHGEN} we show the relative contributions from 
the 1st, 2nd, and 3rd terms in Equation~\ref{E:ENSTRGEN} to the enstrophy generation
(blue, green and red lines).   
Note that these terms represent enstrophy generation rates and therefore their dimension 
is enstrophy over time.
For comparison, we also show the enstrophy (black solid line) normalized to its value at the injection
(also shown in Figure~\ref{F:ENSTROPHYV18}). 
It can be seen clearly from this figure that the compression and the baroclinic terms 
(2nd and 3rd terms in Equation~\ref{E:ENSTRGEN}) are fluctuating around zero,
thus their time integral will be negligible and they do not contribute significantly to the enstrophy. 
The stretching term (1st term in Equation~\ref{E:ENSTRGEN}) 
has a shape similar to the of the enstrophy (black solid line).
As expected, this term is always positive \citep[e.g.,][]{BuxtonGanap2010}, 
and, as a consequence, the contribution from stretching dominates the enstrophy generation.
Note, however, that \cite{BuxtonGanap2010} used equations of incompressible fluid dynamics
in their analysis, and our simulations were based on compressible fluid dynamics, which is more 
appropriate for astrophysical applications.

\subsection{Applications to 3C 75}
\label{SS:RESULTS3C75}

The two bipolar jet system in 3C 75 is located in the dumbbell shaped twin Wide Angle Tail (WAT) 
radio galaxy NGC 1128 at the center of the nearby galaxy cluster Abell 400 at a redshift of 0.023 
\citep[e.g.,][]{OwenET1985,HudsonET06}. 
\cite{HudsonET06} concluded that the AGNs in 3C 75 form a bound system originating from a 
previous merger, and they are both contained in a low entropy core moving through 
the intracluster medium at a relative speed of 1200 km/sec. 
In Figure~\ref{F:IMAGE3C75} the radio and optical (SDSS) images of \THREEC are shown 
with green and red colors \citep{HardcastleSakelliou2004}.
The two AGNs, the sources of the jets, at the center of the image, 
are located at about 7 kpc from each other in projection.

3C 75 is one of the best sources to study the different physical mechanisms capable 
of bending jets
(e.g., ram pressure due to the ambient gas as the jets move, 
collision with dense gas in the ambient medium, and collision with other jets).
The ram pressure from a medium in transverse relative motion to the jet is the most 
common reason why jets show large-scale bends. 
The smooth changes in the direction of the jets in 3C 75 farther from their AGNs
are due to the relative movement of the AGNs to the ambient medium.
The direction of the bending of the jets (towards north-east) 
suggests that 3C 75 is moving to the south-west. 
The eastern jet of the northern AGN seems to bend suddenly to the north 
about 25 kpc east of the core, where an enhancement can be seen in the X-ray 
emission \citep{HudsonET06}, perhaps because of an encounter with a denser 
gas cloud in the intergalactic medium, or with the atmosphere of a galaxy.
However, the western jet from the northern AGN, 
after traveling about 15 kpc from its source,
bends to the north by about 45\DEG where its path crosses that of the 
northern jet of the southern AGN.
No X-ray enhancement associated with extra gas can be seen in this location and
the southern AGN seems to be propagating towards the north-west
without changing its direction \citep{HudsonET06}.
This morphology suggests that the bending of the jet from the 
northern AGN at this location is a result of a collision with the jet from the southern AGN.

\begin{figure}[t]
\includegraphics[width=.478\textwidth]{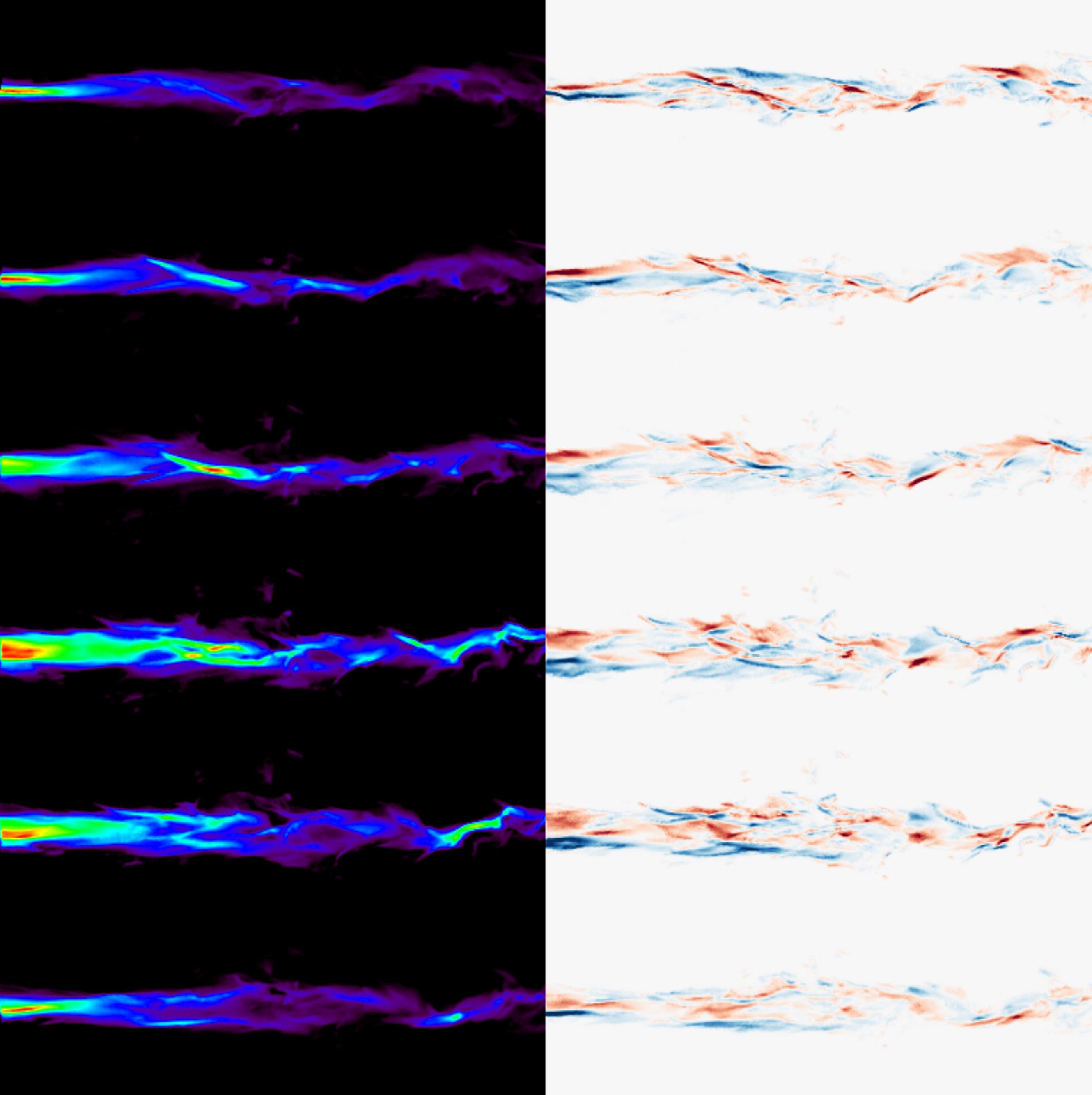}
\caption{
    Projections of the total energy and helicity (left and right panels) 
    of the two jets just after the collision 
    using different rotation angles around the direction of the jet velocity (left to right):
    $\varphi =$ 0\DEG, 30\DEG, 60\DEG, 90\DEG, 120\DEG, 150\DEG\ (top to bottom)
    from our hydrodynamical simulations with jet injection velocities of
    v$_1$ = 18000 \KMSEC and v$_2$ = 10000 \KMSEC, and an impact parameter of P = 0.3. 
    Blue and red colors represent positive and negative helicities.
\vspace{.3 cm}
\label{F:ROTENEHELI}
}
\end{figure}

In Figure~\ref{F:ROTENEHELI} we show projections of the
total energy and helicity along the jet in the interaction region (left and right panels)
from our simulation with v$_1$ = 18000 \KMSEC and v$_2$ = 10000 \KMSEC, 
and an impact parameter of P = 0.3 (model P030V10) 
using different rotation angles around the direction of the propagation of the jet:
$\varphi =$ 0\DEG, 30\DEG, 60\DEG, 90\DEG, 120\DEG, 150\DEG\ (top to bottom).
Blue, red, and white colors represent positive, negative, and zero helicities.
In projection, a double helical feature can be seen in some viewing angles
(especially at $\varphi =$ 0\DEG), which 
may give the impression that the two jets are spiraling around each other.
However, our simulation shows that, in this case the jets merge,
the faster jet takes over the slower one, and breaks into filaments due to
the enhanced instabilities arising from the collision.
The two filaments travel with a speed close to that of the faster jet and have no 
coherent rotational velocities perpendicular to their propagation, 
which would be necessary for helical motion.
However, the filaments seem to have different helicities, coherent on the scale of a few kpc 
along their propagation direction, which may provide stability to these oscillating filaments
(right column in Figure~\ref{F:ROTENEHELI}).

Our hydrodynamical simulations suggest a possible physical explanation for the morphology of \THREEC.
The slower northern jet from the northern AGN collides with a small but finite impact parameter 
with the faster jet from the southern AGN traveling towards north-west, the jets merge, 
and the faster jet takes over the slower jet (most of the gas from the slower jet is grabbed by the faster jet).
The helical-looking morphology is a projection effect: the faster jet 
breaks into filaments due to the enhanced instabilities caused by the collision
with the slower jet. This explanation is consistent with all observed features of \THREEC.

\section{Conclusion}
\label{S:CONCLUSION}

We carried out hydrodynamical simulations to study extragalactic jet collisions.
We found that colliding jets can be cast into two categories: bouncing and merging jets.
We have shown that two fast jets colliding with non-zero impact parameter bounce off 
each other keeping their identities, but jets with very different velocities colliding with 
a small impact parameter merge into one jet, and the faster jet takes over the slower jet.
We have found that the collision enhances the instabilities of the jets;
kpc scale oscillations are generated, and the jets may break up into filaments. 
In some projections, the filaments may show a twisted structure.

In general, magnetic fields in jets can reduce the growth of the Kelvin-Helmholtz
instabilities, and, as a consequence, generate less turbulence, and can also 
affect the transport processes within the jets \citep[e.g.,][]{Hardee2004,Ferrari1998}.
However, the collisions between fast jets included in our simulations 
are so energetic that we expect that the weak magnetic fields 
inferred from observations would affect neither the dynamics of the 
collisions significantly, nor our qualitative results on bouncing and merging of jets.

Our hydrodynamical simulations suggest a physical explanation for the twisted 
radio morphology of 3C 75: strong instabilities are generated in the faster jet by the 
collision with a much slower jet with a small, but finite impact parameter. 
We leave a more quantitative analysis of 3C 75 for a future study.

\acknowledgements
We thank the anonymous referee for detailed comments and suggestions,
which helped to improve the presentation of our results.
The code \FLASH\ used in this work was in part developed by the
DOE-supported ASC/Alliance Center for Astrophysical Thermonuclear
Flashes at the University of Chicago.  
This research has made use of the NASA/IPAC
Extragalactic Database (NED) which is operated by the Jet Propulsion
Laboratory, California Institute of Technology, under contract with
the National Aeronautics and Space Administration.

\clearpage
%
%
\bibliographystyle{apj}

\begin{thebibliography}{99}



\bibitem[Birkinshaw(1991)]{Birkinshaw1991}
 Birkinshaw, M.\ 1991, \mnras, 252, 505 

\bibitem[Buxton \& Ganapathisubramani(2010)]{BuxtonGanap2010}
 Buxton, O., R., H. \& Ganapathisubramani, B., 2010, J. Fluid Mech., 651, 483

\bibitem[Colella(1990)]{Colella1990}
 Colella, P. 1990, J. Comput. Phys., 87, 171
 
\bibitem[Colella \& Woodward(1984)]{CW1984}
 Colella, P., \& Woodward, P. R. 1984, J. Comput. Phys., 54, 174

\bibitem[Edge(2001)]{Edge2001}
 Edge, A.~C.\ 2001, \mnras, 328, 762 

\bibitem[Fabian(1994)]{Fabian94}
 Fabian A. C., 1994, \araa, 32, 277

\bibitem[Fabian(2012)]{Fabian2012ARAA}
 Fabian, A.~C.\ 2012, \araa, 50, 455 

\bibitem[Falceta-Gon{\c c}alves et al.(2010)]{FalcetaET2010}
 Falceta-Gon{\c c}alves, D., de Gouveia Dal Pino, E.~M., Gallagher, J.~S., \& Lazarian, A.\ 2010, \apjl, 708, L57 

\bibitem[Feretti et al.(1999)]{FerettiET1999}
 Feretti, L., Perley, R., Giovannini, G., \& Andernach, H.\ 1999, \aap, 341, 29 

\bibitem[Ferrari(1998)]{Ferrari1998}
 Ferrari, A.\ 1998, \araa, 36, 539 

\bibitem[Fryxell et al.(2000)]{FryxellET2000}
 Fryxell, B., Olson, K., Ricker, P., et al., 2000, \apjs, 131, 273 


\bibitem[Gaspari et al.(2011)]{GaspariET2011}
 Gaspari, M., Melioli, C., Brighenti, F., \& D'Ercole, A.\ 2011, \mnras, 411, 349 

\bibitem[{{Gaspari} {et~al.}(2013){Gaspari}, {Ruszkowski}, \& {Oh}}]{GaspariET2013}
 {Gaspari}, M., {Ruszkowski}, M., \& {Oh}, S.~P. 2013, \mnras, 432, 3401

\bibitem[Gaspari et al.(2012)]{GaspariET2012}
 Gaspari, M., Ruszkowski, M., \& Sharma, P.\ 2012, \apj, 746, 94 

\bibitem[Hardcastle \& Sakelliou(2004)]{HardcastleSakelliou2004}
 Hardcastle, M.~J., \& Sakelliou, I.\ 2004, \mnras, 349, 560 

\bibitem[Hardee(2004)]{Hardee2004}
 Hardee, P.~E.\ 2004, \apss, 293, 117 

\bibitem[Hudson et al.(2006)]{HudsonET06}
 Hudson, D. S.,  Reiprich, T. H. Clarke, T. E. \& Sarazin, C. L. 2006,\  A\&A 453, 433

\bibitem[Laing \& Bridle(2014)]{LaingBridle2014MNRAS}
 Laing, R.~A., \& Bridle, A.~H.\ 2014, \mnras, 437, 3405 

\bibitem[Laing et al.(1999)]{LaingET1999}
 Laing, R.~A., Parma, P., de Ruiter, H.~R., \& Fanti, R.\ 1999, \mnras, 306, 513 

\bibitem[Li \& Bryan(2014a)]{LiBryan2014a}
 Li, Y., \& Bryan, G.~L.\ 2014a, \apj, 789, 54 

\bibitem[Li \& Bryan(2014b)]{LiBryan2014b}
 Li, Y., \& Bryan, G.~L.\ 2014b, \apj, 789, 153

\bibitem[Li et al.(2015)]{LiET2015}
 Li, Y., Bryan, G.~L., Ruszkowski, M., et al.\ 2015, \apj, 811, 73 

\bibitem[Martizzi et al.(2012)]{MartizziET2012}
 Martizzi, D., Teyssier, R., \& Moore, B.\ 2012, \mnras, 420, 2859 

\bibitem[McDonald, Veilleux \& Mushotzky(2011)]{McDonaldET2011}
 McDonald, M., Veilleux, S., \& Mushotzky, R.\ 2011, \apj, 731, 33 

\bibitem[{{McNamara} \& {Nulsen}(2007)}]{McNamaraNulsen2007}
{McNamara}, B.~R., \& {Nulsen}, P.~E.~J. 2007, \araa, 45, 117

\bibitem[Mignone \& McKinney(2007)]{MignoneMcKinney2007MNRAS}
 Mignone, A., \& McKinney, J.~C.\ 2007, \mnras, 378, 1118 

\bibitem[{{O'Dea} {et~al.}(2008){O'Dea}, {Baum}, {Privon}, {Noel-Storr},
               {Quillen}, {Zufelt}, {Park}, {Edge}, {Russell}, {Fabian}, {Donahue},
               {Sarazin}, {McNamara}, {Bregman}, \& {Egami}}]{ODeaET2008}
 {O'Dea}, C.~P., {Baum}, S.~A., {Privon}, G., {et~al.} 2008, \apj, 681, 1035

\bibitem[Owen et al.(1985)]{OwenET1985}
 Owen, F.~N., O'Dea, C.~P., Inoue, M., \& Eilek, J.~A.\ 1985, \apjl, 294, L85 

\bibitem[{{Peterson} {et~al.}(2003){Peterson}, {Kahn}, {Paerels}, {Kaastra},
               {Tamura}, {Bleeker}, {Ferrigno}, \& {Jernigan}}]{PetersonET2003}
 {Peterson}, J.~R., {Kahn}, S.~M., {Paerels}, F.~B.~S., {et~al.} 2003, \apj, 590, 207

\bibitem[{{Peterson} \& {Fabian}(2006)}]{PetersonFabian2006}
 {Peterson}, J.~R., \& {Fabian}, A.~C. 2006, \physrep, 427, 1
 
\bibitem[{{Pizzolato} \& {Soker}(2005)}]{PizzolatoSoker2005}
 {Pizzolato}, F., \& {Soker}, N. 2005, \apj, 632, 821

 \bibitem[Porter et al.(2015)]{PorterET2015}
 Porter, D.~H., Jones, T.~W., \& Ryu, D.\ 2015, \apj, 810, 93 

\bibitem[Pudritz et al.(2012)]{PudritzET2012}
 Pudritz, R.~E., Hardcastle, M.~J., \& Gabuzda, D.~C.\ 2012, \ssr, 169, 27 

\bibitem[Roe(1981)]{Roe1981}
 Roe, P. L. 1981, J. Comput. Phys., 43, 357
 
\bibitem[Salom{\'e} \& Combes(2003)]{SalomeCombes2003}
 Salom{\'e}, P., \& Combes, F.\ 2003, \aap, 412, 657 

\bibitem[Schive(2012)]{Schive2012}
 Schive, H.-Y., Zhang, U.-H., \& Chiueh, T. 2012, IJHPCA, 26, 367

\bibitem[Schive et al.(2010)]{SchiveET2010ApJS186}
 Schive, H.-Y., Tsai, Y.-C., \& Chiueh, T.\ 2010, \apjs, 186, 457 

\bibitem[Tamura et al.(2003)]{TamuraET2003}
Tamura T., Kaastra J. S., Makishima K., \& Takahashi I., 2003, A\&A, 399, 497

\bibitem[Trac \& Pen(2003)]{TracPen2003}
 Trac, H., \& Pen, U.-L.\ 2003, \pasp, 115, 303 

\bibitem[van Loo et al.(2015)]{LooET2015}
 Van Loo, S., Tan, J.~C., \& Falle, S.~A.~E.~G.\ 2015, \apjl, 800, L11 

\bibitem[Yang \& Reynolds(2016)]{YangReynolds2016ApJ818}
 Yang, H.-Y.~K., \& Reynolds, C.~S.\ 2016, \apj, 818, 181 






\end{thebibliography}

%
%
\begin{deluxetable*}{ccccccccccc}[t]
\tablecolumns{9}
\tablecaption{                       \label{T:TABLE1} 
 IDs and the grid of input parameters for different models used in our hydrodynamical simulations.
} 
\tablewidth{0pt} 
\tablehead{ 
 \multicolumn{1}{c}   {v$_{1}$\tablenotemark{a}}          &
 \multicolumn{1}{c}   {v$_{2}$\tablenotemark{b}}          &
 \multicolumn{1}{c}   {M$_{\rm jet}$\tablenotemark{c}}          &
 \multicolumn{1}{c}     {P = 0\,\tablenotemark{d}}          &
 \multicolumn{1}{c}{P = 0.15\,\tablenotemark{d}}          &
 \multicolumn{1}{c}{P = 0.30\,\tablenotemark{d}}          &
 \multicolumn{1}{c}{P = 0.45\,\tablenotemark{d}}          &
 \multicolumn{1}{c}{P = 0.60\,\tablenotemark{d}}          &
 \multicolumn{1}{c}{P = 0.80\,\tablenotemark{d}}          &
 \multicolumn{1}{c}     {P = 1\,\tablenotemark{d}}          &
 \multicolumn{1}{c}{P = $\infty$\,\tablenotemark{e}}
 }
 \startdata  
 18   &   18   &   7.9   &  P000V18  &   P015V18    &   P030V18     &   P045V18   &  P060V18  &  P080V18  &  P100V18  &   SNGV18 \\ \hline
 18   &   16   &   7.0   &  P000V16  &   P015V16    &   P030V16     &   P045V16   &  P060V16  &  P080V16  &  P100V16  &   SNGV16 \\ \hline
 18   &   14   &   6.1   &  P000V14  &   P015V14    &   P030V14     &   P045V14   &  P060V14  &  P080V14  &  P100V14  &   SNGV14 \\ \hline
 18   &   12   &   5.3   &  P000V12  &   P015V12    &   P030V12     &   P045V12   &  P060V12  &  P080V12  &  P100V12  &   SNGV12 \\ \hline
  -     &   10   &   4.4   &  P000V10  &   P015V10    &   P030V10     &   P045V10   &  P060V10  &  P080V10  &  P100V10  &   SNGV10
\enddata
\tablecomments{The IDs indicate the impact parameters and velocities of the second jet.}
\tablenotetext{a}{Injection velocity of the 1st jet in 1000 \KMSEC (fixed).}
\tablenotetext{b}{Injection velocity of the 2nd jet in 1000 \KMSEC.}
\tablenotetext{c}{Jet Mach number of the 2nd jet.}
\tablenotetext{d}{Impact parameters, P, in units of the jet radius (see Section~\ref{SS:InitialConditions}).}
\tablenotetext{e}{Impact parameter of P = $\infty$ refers to our single jet models with different injection velocities (v$_{2}$).}
\end{deluxetable*}  

%
%
%
%

\clearpage 

%
%

%
%
\begin{amssidewaysfigure}
\centering
  \includegraphics[width=0.163\textwidth]{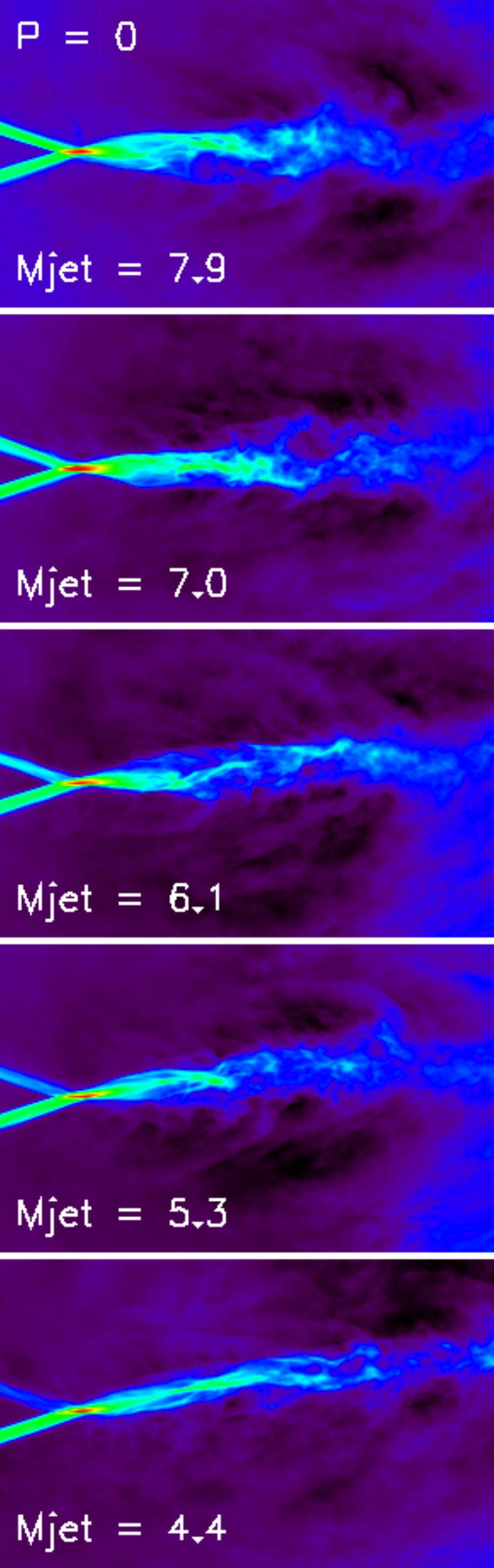}  
  \includegraphics[width=0.163\textwidth]{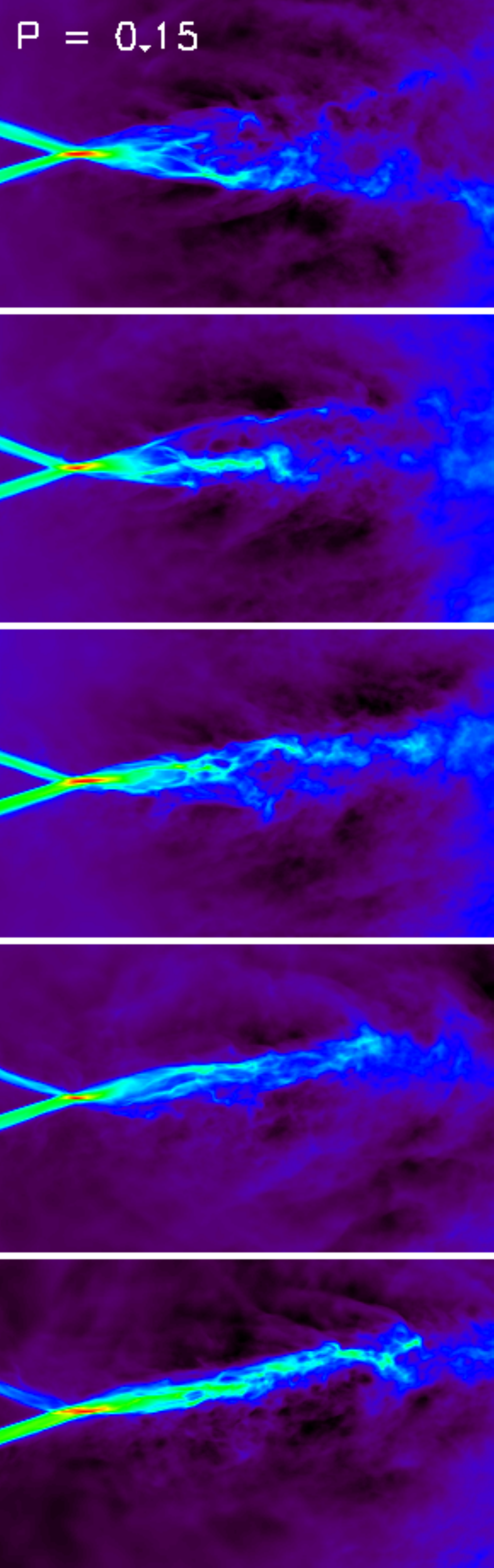}  
  \includegraphics[width=0.163\textwidth]{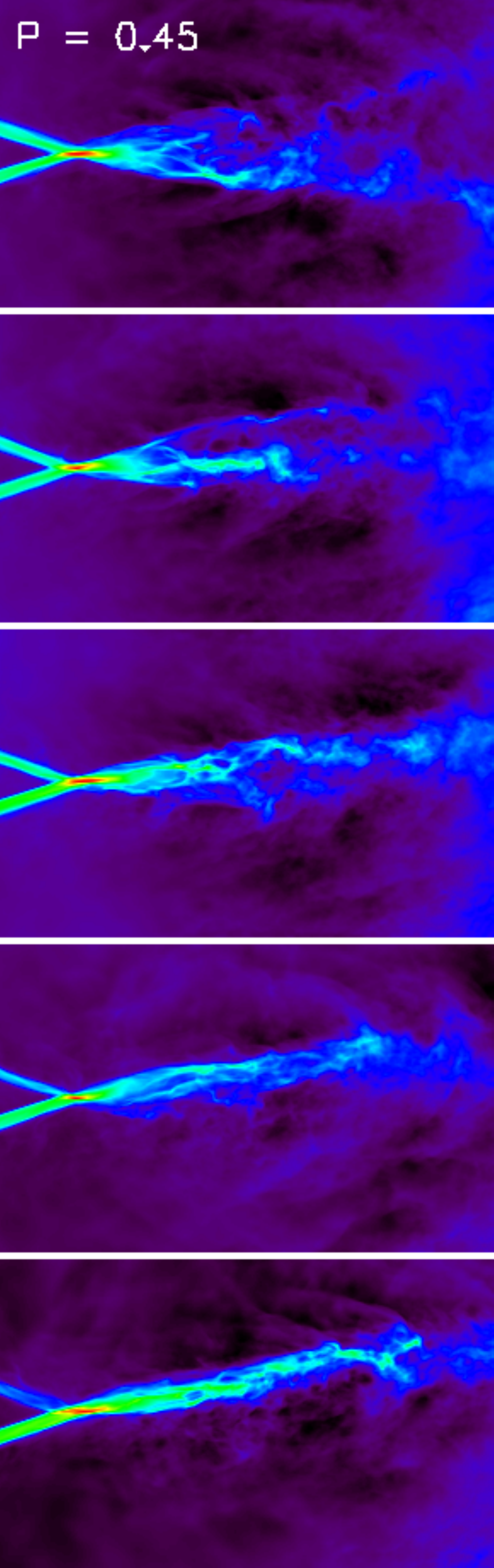}  
  \includegraphics[width=0.163\textwidth]{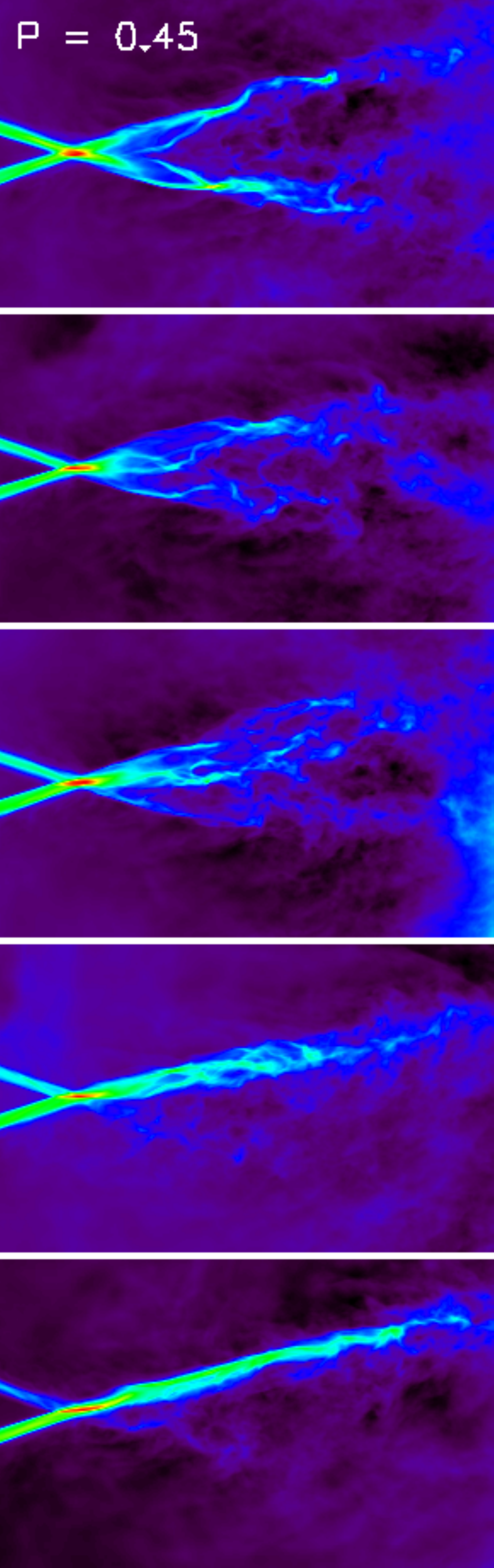}  
  \includegraphics[width=0.163\textwidth]{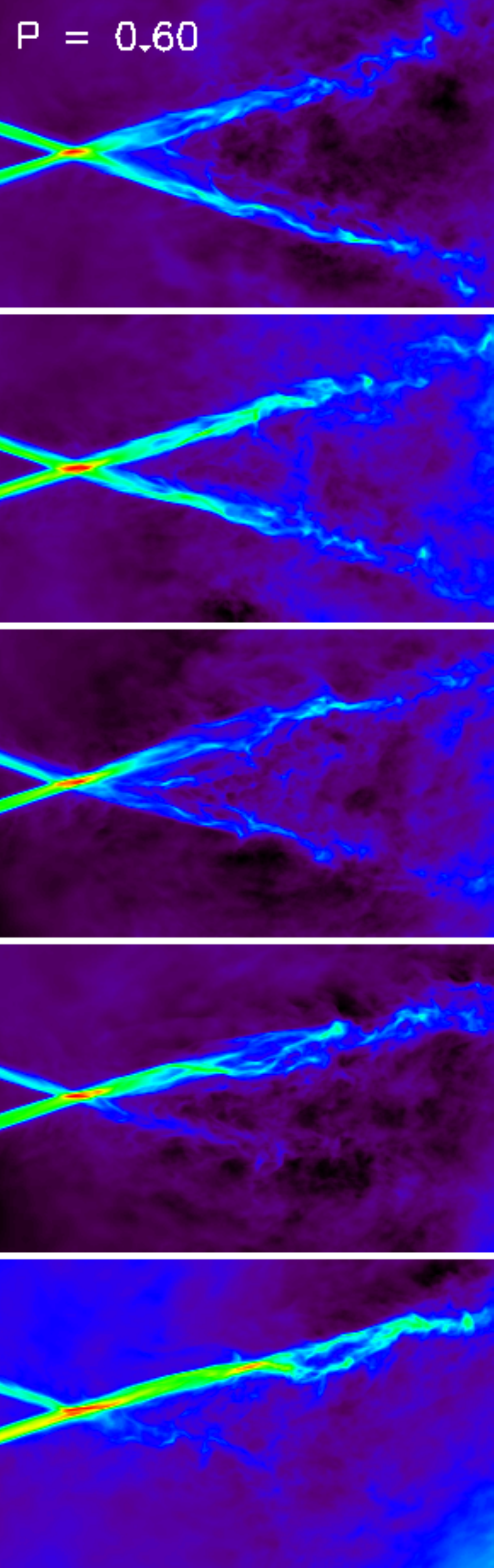}  
  \includegraphics[width=0.163\textwidth]{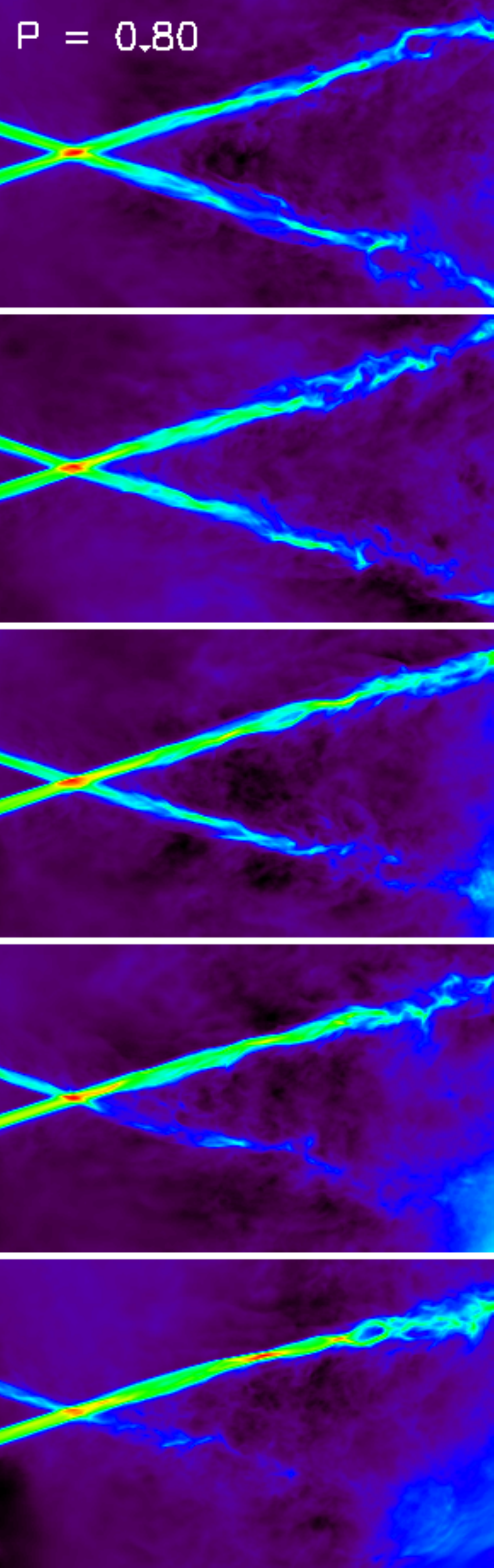}   
\caption{ \protect
 Total energy density (log scale) of colliding jets from our simulations projected to the $(x, y)$ 
 plane with different impact parameters and injection velocities of the second jet.
 Columns relate to distinct impact parameters, P (in units of the diameter of the jet) 
 given at the top of each column, and rows relate to distinct values of 
 the jet internal Mach number, $M_{jet}$.
 The faster jet is propagating from the lower left to the upper right with a fixed Mach number of 7.9.
 See Section~\ref{SS:InitialConditions} for the details of the initial set up of our simulations.
 The red region located on the left hand side of each panel marks 
 enhanced total energy due to the collision of the two jets.
\label{F:PROJZENERALL}
}
\end{amssidewaysfigure}

\newpage

%
%
\begin{evenamssidewaysfigure}
\centering
  \includegraphics[width=0.14\textwidth]{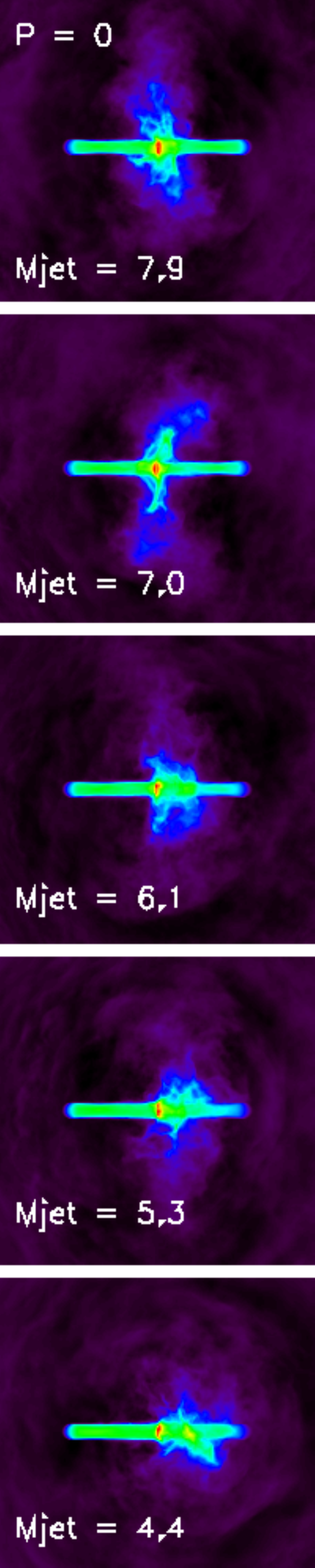}  
  \includegraphics[width=0.14\textwidth]{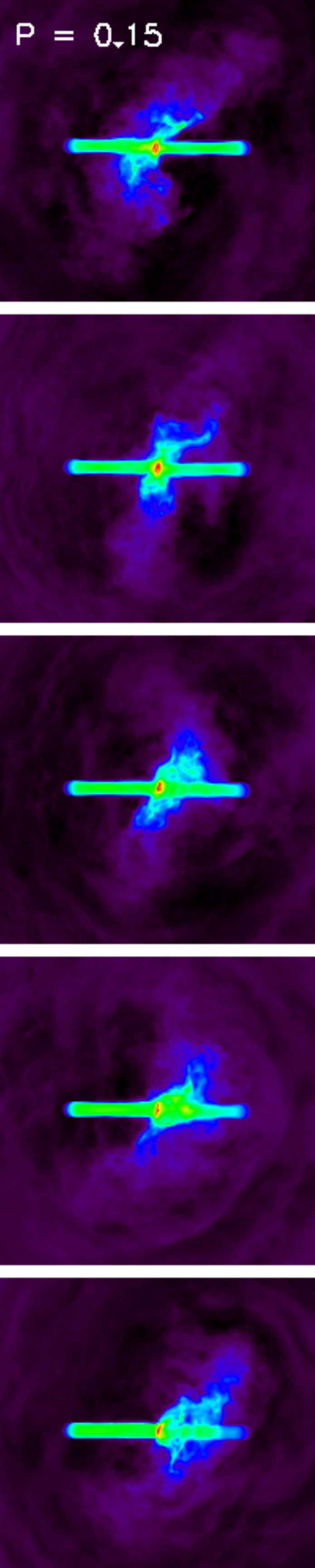}  
  \includegraphics[width=0.14\textwidth]{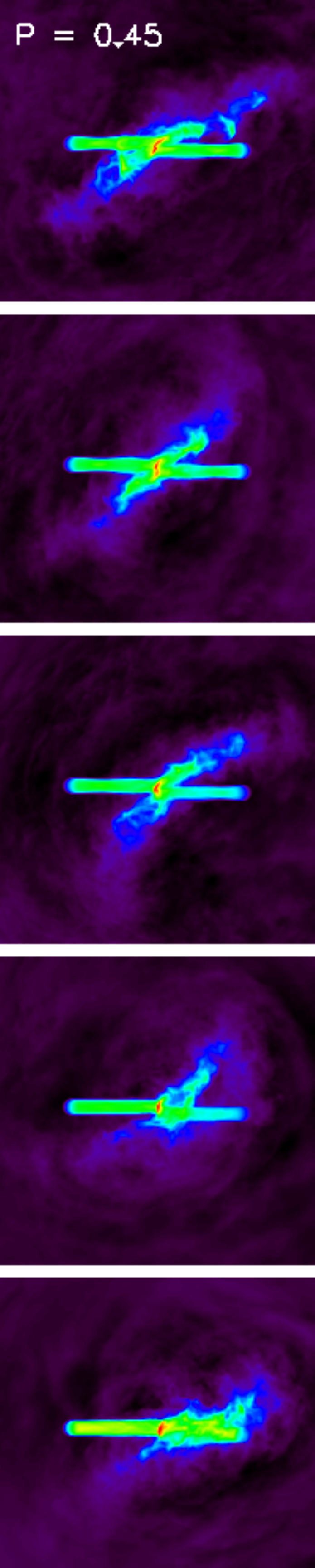}  
  \includegraphics[width=0.14\textwidth]{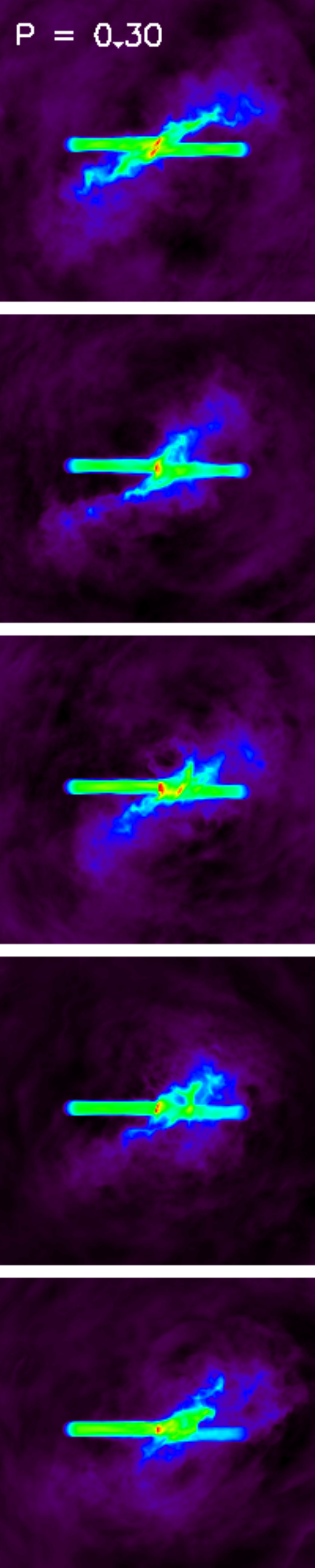}  
  \includegraphics[width=0.14\textwidth]{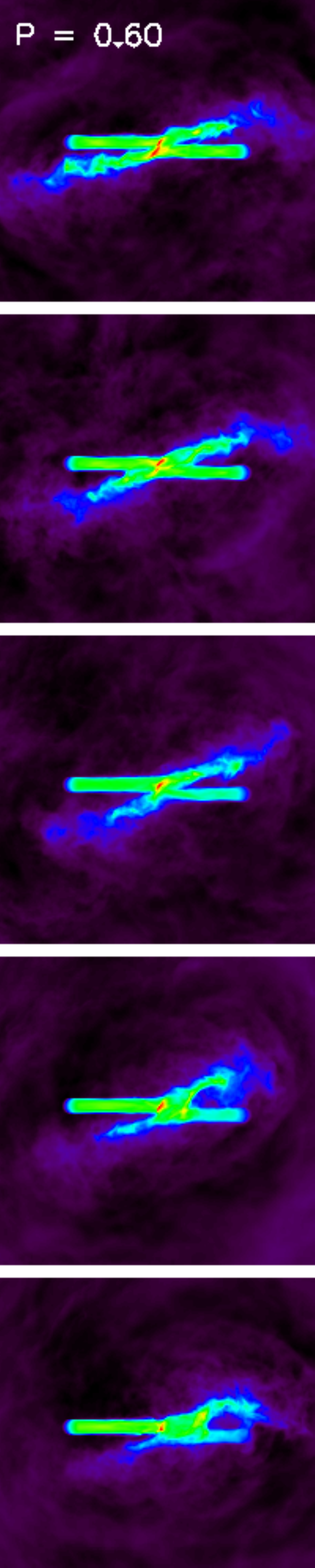}  
  \includegraphics[width=0.14\textwidth]{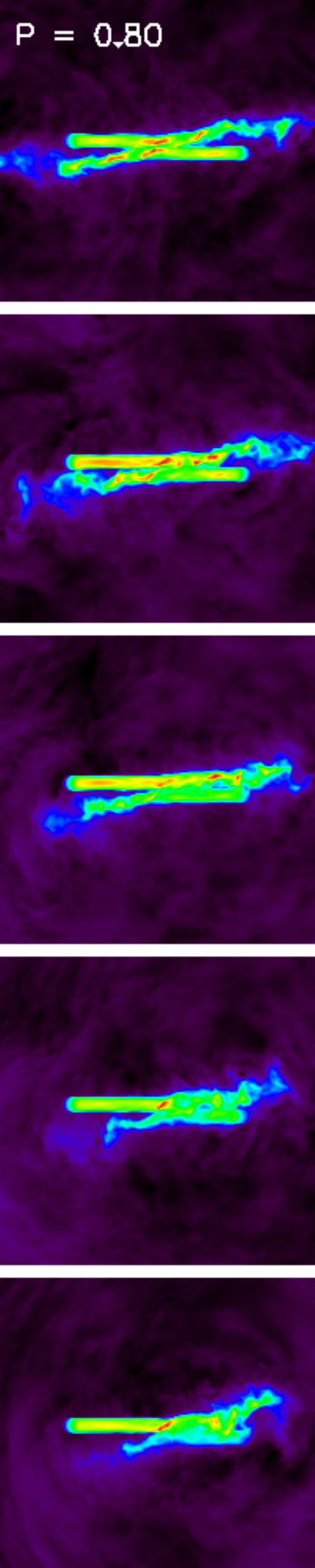}  
\caption{ \protect
 Same as Figure~\ref{F:PROJZENERALL}, but in projection to the $(x, z)$ plane. 
 The faster jet moves from left to right along the $x$ axis with a fixed injection velocity, 
 the second jet movies from right to left with different velocities. 
 The collision region in the middle of each panel is marked with red.
\label{F:PROJYENERALL}
}
\end{evenamssidewaysfigure}

%
%

%
%

%
%

%
%

%
%


\end{document}